\documentclass[aps,twocolumn,nofootinbib,preprintnumbers]{revtex4}
\usepackage{graphicx, float}
\usepackage{hyperref}
\usepackage{amsmath}
\usepackage{amssymb}
\usepackage[usenames,dvipsnames]{xcolor}

\newcommand{\cc}{{\rm c.c.}}
\newcommand{\cL}{{\cal L}}

\newcommand{\beq}{\begin{eqnarray}}
\newcommand{\eeq}{\end{eqnarray}}

\def\simlt{\stackrel{<}{{}_\sim}}
\def\simgt{\stackrel{>}{{}_\sim}}

\def\mone{m_{\tilde{t}_1}}
\def\mtwo{m_{\tilde{t}_2}}

\newcommand{\leqn}[1]{(\ref{#1})}

\begin{document}

\preprint{CERN-PH-TH-2015-217}
\preprint{ACFI-T15-15}

\title{Stop-Catalyzed Baryogenesis Beyond the MSSM}

\author{Andrey Katz}
\affiliation{Theory Division, CERN, CH-1211 Geneva 23, Switzerland}
\affiliation{Universit\'{e} de Gen\`{e}ve, Department of Theoretical
  Physics and Center for Astroparticle Physics (CAP),  
24 quai E. Ansermet, CH-1211, Geneva 4, Switzerland}
\author{Maxim Perelstein}
\affiliation{Laboratory for Elementary Particle Physics, Cornell
  University, Ithaca, NY 14853, USA} 
\author{Michael J. Ramsey-Musolf}
\affiliation{Amherst Center for Fundamental Interactions, Department
  of Physics, 
University of Massachusetts-Amherst Amherst, MA 01003, USA}
\affiliation{Kellogg Radiation Laboratory, California Institute of Technology, Pasadena, CA 91125 USA} 
\author{Peter Winslow}
\affiliation{Amherst Center for Fundamental Interactions, Department
  of Physics, 
University of Massachusetts-Amherst Amherst, MA 01003, USA}

\begin{abstract}
\noindent 
Non-minimal supersymmetric models that predict a tree-level Higgs
mass above the Minimal Supersymmetric Standard Model (MSSM) bound are
well motivated by naturalness 
considerations. Indirect constraints on the stop
sector parameters of such models are significantly relaxed compared to the MSSM; in
particular, both stops can have weak-scale masses. We revisit the
stop-catalyzed electroweak baryogenesis (EWB) scenario in this
context. We find that the LHC measurements of the Higgs boson
production and decay rates already rule out the possibility of
stop-catalyzed EWB. We also introduce a gauge-invariant analysis framework that may
generalize to other scenarios in which interactions outside the gauge sector drive the electroweak phase transition.    

\end{abstract}

\maketitle

\section{Introduction}

The origin of matter-antimatter asymmetry in the Universe is a
longstanding problem at the interface of particle  and nuclear physics
with cosmology. This issue cannot be 
addressed within the Standard Model (SM) and requires physics beyond the
SM. One of the most interesting possibilities, which has attracted much
attention in recent years, is the electroweak baryogenesis (EWB) scenario
(see~\cite{Morrissey:2012db} for review),
where the baryon asymmetry is produced during the electroweak phase transition
(EWPT). This mechanism requires new
physics beyond the Standard Model (BSM)  at the weak scale ($\sim
100$~GeV) for two different reasons:   

\begin{itemize}

\item The SM with the Higgs boson mass above $\sim 80$~GeV does not satisfy the
  Sakharov criterion~\cite{Sakharov:1967dj} of departure from thermal
  equilibrium because the EWPT is a 
  cross-over, rather than a strong 1st-order
  transition~\cite{Gurtler:1997hr,Laine:1998jb,Csikor:1998eu,Aoki:1999fi}. In
  order for   EWB to be viable, one inevitably needs new bosonic fields that
 couple to the Higgs and significantly change the EWPT dynamics.

\item Although the SM does violate CP symmetry, the effects are highly
  suppressed at temperatures $T\sim 100$ GeV by quark Yukawa
  couplings. Consequently, even if a SM universe admitted a strong 1st-order
  EWPT the produced 
  baryon asymmetry is too
  small~\cite{Gavela:1993ts,Huet:1994jb,Gavela:1994dt}. New sources of
  CP violation at 
  the weak scale are required.   

\end{itemize} 

Weak-scale supersymmetry (SUSY) is a well-motivated extension of the
SM which can address both these problems.  
Electroweak gaugino/Higgsino~\cite{Li:2008ez,Cirigliano:2009yd} and
scalar~\cite{Kozaczuk:2012xv} phases can provide the new sources of
CP violation (CPV), while the nature of EWPT
can drastically change either in the presence of low-mass
stops~\cite{Carena:1996wj,Delepine:1996vn,Carena:2008vj}, or 
in models with extended Higgs sectors, e.g. additional singlet in the
Next-to-Minimal Supersymmetric Standard Model
(NMSSM)~\cite{Pietroni:1992in,Menon:2004wv,Huber:2006wf}. In both cases, present and 
future experimental probes 
of BSM physics may provide conclusive tests. 
Present limits on the permanent electric dipole moments (EDMs) of the
electron, neutron, and neutral atoms place stringent constraints on
these CP-violating sources, and future EDM searches may probe the
remaining CPV parameter space~\cite{Engel:2013lsa}. At the same time,
searches for new scalar particles at the CERN Large Hadron Collider
may uncover the ingredients needed for a first order EWPT. 

In this paper, we focus on the possibility that low-mass stops may
give rise to the first order EWPT, the the so-called {\it
  stop-catalyzed electroweak 
  baryogenesis} scenario. The basic idea  is that a very light stop 
($m\approx 100$~GeV) modifies the Higgs potential at finite temperatures
via 
quantum loop effects, inducing a barrier along the Higgs direction
between the electroweak symmetric and broken vacua and  
triggering a strongly 1st-order EWPT. It has been 
already shown in~\cite{Curtin:2012aa,Cohen:2012zza} that this scenario
is no longer viable 
in the context of Minimal Supersymmetric Standard Model (MSSM), in
light of the LHC measurements of the Higgs 
boson mass, production cross sections and decay rates (for pre-LHC
works on the same subject see~\cite{Menon:2009mz}). 

In this work we
analyze the stop-catalyzed baryogenesis in a more generic SUSY
framework. Specifically, we consider adding small, hard-SUSY breaking
terms to the Higgs potential. (Such terms can be dynamically
generated at the few-TeV energy scale; for examples,
see~\cite{Batra:2003nj,Maloney:2004rc,Dine:2007xi,Lu:2013cta}). The
primary motivation for   
models of this type comes from considerations of
naturalness~\cite{Katz:2014mba}. The new 
potential terms can give tree-level contributions to the SM-like Higgs
mass, allowing for a SUSY theory with an SM-like Higgs at 125~GeV
{\it independently of the stop sector parameters}. Spectra with two
relatively light stops can then still be viable, reducing
fine-tuning. One may hope that the wide-open stop parameter space of
these models may also allow them to accommodate the stop-catalyzed
EWB scenario. This paper will explore whether this is indeed
the case. 

Unfortunately, we find that even in this broader framework,
there is still no parameter space which is 
compatible with both the strong 1st order EWPT and the LHC Higgs 
measurements. The key observation which leads us to this conclusion is
that the light stop can change the order of the phase transition only
if its coupling to the Higgs is close to its maximal possible value,
which occurs when the light stop is nearly a pure gauge eigenstate
(that is, $\tilde{t}_1\approx \tilde{t}_L$ or $\tilde{t}_1\approx
\tilde{t}_R$). On the other hand, in the small-mixing limit, loops of
the light stop induce very large ($\sim 100$\%) shifts in the Higgs
couplings to photons and gluons, in contradiction with the LHC data
that requires these couplings to be within $10-20$\% of their SM
values.  
It has been proposed in Ref.~\cite{Carena:2012np} that this problem
can be resolved if one assumes an appreciable invisible rate of the
Higgs particle into 
neutralino pairs, which would compensate for the growth in $hgg$
coupling due to the light stop. We show that this modification is also
no longer compatible with the 
LHC Higgs data. Therefore, we conclude that stop-catalyzed EW
baryogenesis is excluded by data over the entire parameter space, even
if the MSSM Higgs mass constraint is removed.  

The paper is organized as follows. In Sec.~\ref{sec:SUSY} we briefly review
the idea behind stop-catalyzed baryogenesis and describe in detail our
framework of SUSY beyond MSSM. In Sec.~\ref{sec:Higgs} we review the
existing experimental constraints on the light stops, including both direct LHC searches and indirect constraints 
from measurements of the Higgs properties. In Sec.~\ref{sec:EWPT} we discuss  the conditions for a strong 1st order EWPT with 
an emphasis on gauge-invariance of the analysis.  The framework that we introduce in this context may generalize to other scenarios in which new interactions outside the gauge sector drive the EWPT dynamics.
Sec.~\ref{sec:results} discusses some other
technical details of our analysis, as well as its main results. Finally, the conclusions are summarized in Sec.~\ref{sec:conc}.

\section{SUSY baryogenesis and SUSY beyond the MSSM}
\label{sec:SUSY}

In the SM with the Higgs mass above $\gtrsim 80$~GeV the EWPT is a
cross-over \cite{Gurtler:1997hr,Laine:1998jb,Csikor:1998eu,Aoki:1999fi}. 
Assuming that the phase transition occurs in a single step, any viable EWB scenario, thus,  requires augmenting the Higgs 
sector with new bosonic degrees of freedom that
couple strongly to the Higgs (for alternatives to the single-step phase transition 
scenario, see, {\em e.g.}
Refs.~\cite{Profumo:2007wc,Patel:2012pi,Patel:2013zla,Profumo:2014opa,Jiang:2015cwa,Blinov:2015sna,Inoue:2015pza}). One     
generic possibility is new scalar particles that generate 
a barrier between the electroweak symmetric and broken vacua in the
Higgs thermal potential via quantum loops, leading to 
a strongly 1st order EWPT. SUSY provides a natural candidate for such a
particle: the stop. The coupling of the stop to the Higgs is predicted
by SUSY and it is equal (up to the trilinear
term) to the top Yukawa squared, an order-one number. It has been shown
in~\cite{Carena:1996wj, Carena:2008vj,Carena:2012np}
that light stops, with masses roughly in the 
$100\ldots 120$~GeV range, can trigger a strongly 1st order EWPT, if
the stops are not heavily mixed and the $\tan \beta$ is large, $\gtrsim 10$. 
The reason for this
constraint is that mixing reduces the effective coupling between the 
stop and the  
SM-like Higgs. Most of the work on this scenario assumed that the
light stop is purely right-handed, to avoid introducing a very light
sbottom. In this paper, we do not make this assumption, considering
instead the most general stop sector characterized by three
parameters, the physical stop masses $\mone$ and $\mtwo$, and the
rotation angle from gauge to mass eigenbasis, $\theta_t$. (In some of
the plots, we will find it useful to trade $\theta_t$ for the mixing
parameter $X_t$, defined as $X_t \equiv M^2_{LR}/m_t$.) Doing so will
allow 
us to quantify the extent of mixing for which the strong 1st-order transition
is possible, as well as to contrast the parameter space required by
the EWB scenario with that allowed by the LHC data.

Discovery of the 125 GeV Higgs disfavors the pure
MSSM, which predicts $m_h \leq m_Z$ at tree level. The one-loop
contribution to the Higgs mass-squared is given by  
\beq
\Delta m_h^2 = \frac{3 y_t^2}{4\pi^2} \cos^2 \alpha\ m_t^2 \log\left(
  \frac{m_{\tilde t_1} m_{\tilde t_2}}{m_t^2}
\right) + \ldots
\eeq
where the ellipsis stand for the terms proportional to the Higgs
mixing and higher-loop
corrections~\cite{Heinemeyer:1998kz,Heinemeyer:1998np,Carena:2000dp}. Very
high stop masses are 
required to accommodate the measured Higgs mass. For example, if
$m_{\tilde t_1} \approx 100$~GeV, as required in the stop-catalyzed
EWB scenario, the second stop 
would need to have a mass of order 100~TeV to~PeV, unless $X_t \gtrsim
1$~TeV~\cite{Draper:2011aa}. 
Such a spectrum
would imply significant fine-tuning in the electroweak scale. In addition,
the hierarchy of three or more orders of magnitude between the
left-handed and right-handed squark soft masses would be challenging
to explain from model-building point of view. Finally, 
even if these issues are ignored, the purely-MSSM version of the
stop-catalyzed EWB is now in direct conflict with data, since the
light stops loops give unacceptably large shifts to the $hgg$ and
$h\gamma\gamma$ couplings~\cite{Curtin:2012aa,Cohen:2012zza}.

In this paper we take a different approach, following
Ref.~\cite{Katz:2014mba}. We assume that additional TeV-scale physics 
beyond the MSSM provides  
a new tree-level contribution to the Higgs mass, reducing the need for
large radiative corrections. There are many examples of  
such new physics, including (but not limited to) higher-dimensional
F-terms~\cite{Dine:2007xi,Lu:2013cta}, or non-decoupling
D-terms~\cite{Batra:2003nj,Maloney:2004rc}.   
If all non-MSSM states introduced by such models are assumed to be sufficiently heavy,
with masses around 1 TeV or above, their effects at the $\sim 100$~GeV
scale, relevant for both the EWPT and Higgs LHC phenomenology, can be
parametrized as new hard-SUSY breaking terms in the Higgs
potential.\footnote{Some SUSY models that lift the tree-level Higgs
  mass, e.g. nMSSM, introduce new states with masses $\sim
  100$~GeV. These states can have an important effect on the
  EWPT~\cite{Menon:2004wv,Huber:2006wf}. We will not consider such
  models here.} 
These effects can be parametrized by the generic
two-Higgs doublet model (2HDM) potential: 
\beq\label{eq:2HDM}
V &  = & M_u^2 |H_u|^2 + M_d^2 |H_d|^2 + (b H_u \cdot H_d + \cc) + \nonumber\\
&&  \hskip -0.5cm \frac{\lambda_1}{4} |H_u|^4 + \lambda_2 |H_u|^2 (H_u
\cdot H_d + \cc) + \lambda_3 |H_u|^2 |H_d|^2 +\nonumber  \\
&& \hskip -0.5cm\frac{\lambda_4}{2} (H_u \cdot H_d + \cc)^2 + \lambda_5
|H_u \cdot H_d|^2 + \nonumber \\
&& \hskip -0.5cm\lambda_6 |H_d|^2 (H_u \cdot H_d + \cc) + \frac{\lambda_7}{4}|H_d|^4~. 
\eeq 
In the MSSM, the coefficients are given by 
\beq \label{eq:lMSSM}
&&\lambda_1 = \lambda_7 = \frac{g^2 +{g'}^2}{2}, \ \ \lambda_3 =
\frac{g^2 - {g'}^2}{4}, \nonumber \\ && \lambda_5 = - \frac{g^2}{2}, \ \ \lambda_2=\lambda_4=\lambda_6=0.   
\eeq
Additional BSM physics can change these relationships. While such corrections explicitly break
SUSY, one should note that the potential~\leqn{eq:2HDM} is merely an effective description valid at
energy scales $\sim 100$~GeV, where SUSY is already broken. In the
full UV theory, the corrections are generated by integrating out new
physics at energy scales comparable to the soft SUSY-breaking scale, $\sim
1$~TeV.  
 
Not every term in the 2HDM potential can be easily generated by a 
UV-complete supersymmetric theory. For example, it difficult to see how any 
supersymmetric theory can generate non-zero coefficients $\lambda_2$,
$\lambda_4$ and $\lambda_6$ at the tree level.
The other 
coefficients can be generated either by new F-terms or by new D-terms
at multi-TeV scale. However, only $\delta \lambda_1$ can give a substantial
contribution to the SM-like Higgs mass without introducing new light
states.\footnote{It has been shown in Ref.~\cite{Katz:2014mba} that
  $\delta \lambda_5$ alone can also solve the problem of the Higgs
  mass. However, this solution either works for $\tan \beta \lesssim
  3$, where there is no 1st order EWPT, or requires $\delta \lambda_5 >1$,
  which can probably be UV-completed only with new light
  states. Therefore we will disregard this option.} 
For moderately large $\tan \beta$, a tree-level 125 GeV Higgs mass
requires $\delta \lambda_1 \lesssim 
0.1$, which can be easily accommodated in the low-energy effective field theory and
UV-completed at a scale of $2 \ldots 3$~TeV~\cite{Katz:2014mba}. Such
hard-SUSY breaking term will be our basic
assumption. Its presence completely removes the Higgs mass constraint
on the stop masses and mixing. Apart from $\lambda_1$, all other 2HDM coefficients are assumed
to be equal to the MSSM values.

Parenthetically we note that UV-complete theories which induce
a necessary $\delta \lambda_1$ term will often also induce $\delta 
\lambda_5$ and $\delta \lambda_7$ of the same order of magnitude. We
will disregard these terms in our analysis because they have no
significant effect on the SM-like Higgs mass. Moreover, if the 2HDM is
not in the full decoupling regime, these terms usually only make the
fit of the Higgs couplings worse. Therefore the limit where $\delta\lambda_{5,7}\to 0$
can be regarded as the best possible case for the light stop SUSY
scenario.

Two further simplifying assumptions will be made in our
analysis. First, we assume that the 2HDM is in the decoupling regime,
which for all practical purposes means $m_A \gtrsim 800$~GeV. 
In this regime, the additional Higgs
bosons beyond the SM-like 125 GeV state are sufficiently heavy to have
no effect at the EWPT critical temperature, $T_c \sim 100$
GeV.\footnote{We assume in this paper $m_A \approx m_{H} \approx
  m_{H^\pm}$. While this is not true in  generic 2HDM, and an
  appreciable 
  hierarchy between these masses can have peculiar consequences on the
EWPT (see e.g.~\cite{Dorsch:2013wja,Dorsch:2014qja}), it is difficult to
see how this hierarchy can be 
realized in a supersymmetric theory with no additional light states.}
This assumption
is strongly motivated by the agreement of the LHC Higgs measurements
with the SM: any deviation from these limits would only further
strengthen the LHC constraints on the scenario in the large $\tan
\beta$ regime of a supersymmetric theory.
Second, we ignore the effect of all the
superpartners, other than the stops, on the Higgs couplings and its
thermal potential at $T\sim T_c$. This is motivated both by the
non-discovery of superpartners at the LHC, and by the fact that the
Higgs coupling to stops is the strongest among its couplings to
the superpartners. For example, while electroweak gauginos and
Higgsinos may be present at a few-hundred GeV scale (and indeed may be
needed to provide CP-violating phases in the EWB scenario), their
effect on the Higgs properties is subdominant to that of the
stops. With these assumptions, both the EWPT and the collider
constraints depend on just four unknown parameters ($\mone$, $\mtwo$,
$\theta_{\tilde t}$, and $\tan\beta$), enabling us to study the parameter space
comprehensively and draw robust conclusions.      

The spectrum required for the stop-catalyzed EWB would
necessarily require some degree of fine-tuning. At one loop, stop
masses receive a quadratically divergent QCD radiative correction, cut
off by the gluino mass  $m_{\tilde{g}}$. Complete naturalness requires
$m_{\tilde{g}}\simlt 2m_{\tilde{t}}$ in the case of Majorana gluino,
or $m_{\tilde{g}}\simlt 4m_{\tilde{t}}$ if the gluino is
Dirac~\cite{Brust:2011tb}. In either case, if the light stop is close
to 100 GeV, gluinos must occur below 400 GeV; this possibility is
comprehensively ruled out by the LHC data. Given the gluino bounds, a
tuning of ${\cal O}(10-100)$ is required to accommodate such a light
stop. Our philosophy in this paper is to not be concerned about this;
we would like to know whether or not the stop-catalyzed EWB is 
in agreement with the data, regardless of fine-tuning
issues. 

\section{Constraints from Direct Searches and Higgs Measurements}
\label{sec:Higgs}

First, we consider direct searches for stops at the LHC, where the
light stop in the $100-120$ GeV mass range would be copiously
produced. The sensitivity of the searches depends very strongly on the
stop decay channels. There are two possible options, corresponding to 
R-parity conserving and R-parity violating scenarios. In the R-parity
conserving framework a light stop can decay into $\tilde{\chi}^0 b
W^{(*)}$ via an off-shell top (and possibly also an off-shell
$W$). Alternatively, in the mass range where 3-body stop decay is
prohibited, one can expect the decay mode $\tilde t \to c \tilde{\chi}^0$ to
compete with the four-body decay. All these decay channels were studied 
by ATLAS and CMS collaborations. The three- and four-body decays are
searched for in monoleptonic 
channels~\cite{Aad:2014kra,CMS-PAS-SUS-13-011}. While 
most of the parameter space is excluded, the bounds are
discontinuous, and become much weaker or even completely disappear near
the borderlines between three- and four-body decays of the
stop. Moreover, none of these searches is optimized for the
four-body decays, and the constraints in this region are weak. 
This leaves a very light stop in the $\sim 100$~GeV mass range still a
viable possibility. The searches for the stop decays in the two-body
$c\chi^0$ final state~\cite{CMS-PAS-SUS-13-009, Aad:2014nra} are more
decisive and exclude the stops below the mass of $\sim 200$~GeV under the
assumption that the branching ration (BR) into this mode
is~1. However, the exact BRs in this part of parameter space 
depend on the neutralino mixing angles, stop mixture, and possible flavor
violation in the scalar sector beyond MFV.\footnote{It has recently
  been claimed in Ref.~\cite{Batell:2015zla} that the light stop
  scenario can be completely ruled out due to stoponium formation. It would be
  interesting to see whether a dedicated analysis by the experimental
  collaborations confirms this claim.} 

Another possibility is R-parity violation (RPV). In this
case it is conceivable that the light stop is at the bottom of the
spectrum and decays directly into SM states. Of course this possibility is
excluded in the case of lepton-number violation. Searching for light
stops in the 
baryon-number violating scenario, 
where the stop decays into two jets, with or without b-tag, is 
more challenging. However, recently CMS has excluded RPV 
stops below 200~GeV~\cite{Khachatryan:2014lpa}, rendering this option
irrelevant for the stop-catalyzed baryogenesis. Even stronger
constraints have been obtained from an ATLAS boosted
search~\cite{ATLAS-CONF-2015-026}, but they apply only for the
b-tagged scenario. 

To summarize, while direct constrains have already cornered the
possibility of the light stops, the parameter space is not yet
completely closed and the constraints are model-dependent. Therefore,
we will now turn to analyzing indirect constraints, which are more robust.    

An important set of constraints on the light stop scenario comes from the
electroweak precision measurements. Split scalar
multiplets at the electroweak scale, such as stops and bottoms in the presence
of mixing, contribute to the S and T 
parameters~\cite{Drees:1991zk}. But the strongest constraints currently 
come from the measurement of the Higgs properties at the LHC. In
particular, loops of very light stops significantly modify
the coupling of the Higgs to the photons and gluons. If the additional Higgs bosons of the 2HDM are not too heavy, 
Higgs-fermion couplings can also be modified. In the decoupling approximation, which is almost always
true in the SUSY context, and for moderately large $\tan \beta$ this
effect dominantly modifies the Higgs couplings to the 
down-type quarks and the taus~\cite{Blum:2012kn}.

Of course, the latter effect can be easily circumvented simply by
decoupling the heavy Higgses. For example, if the heavy Higgs 
masses are around 800~GeV, we expect $\sim 3\%$ correction to the $h
\to b \bar b$ rate, much too small to be detected with the currently
available data. On the other hand, corrections to the
couplings to photons and gluons are much harder to
address. The stop loop contribution to the $h\gamma \gamma$
and $hgg$ couplings both scale in the small mixing regime
approximately as  
\beq\label{eq:hgg}
g_{hgg}/g_{hgg}^{SM} - 1 \approx \frac14 \left(
  \frac{m_t^2}{m_{\tilde t_1}^2} + \frac{m_t^2}{m_{\tilde t_2}^2} -
  \frac{m_t^2 X_t^2}{m_{\tilde t_1}^2 m_{\tilde t_2}^2} \right).
\eeq      
Without mixing, a $100$ GeV stop produces an ${\cal O}(100\%)$
correction to the couplings, well beyond the $10-20$\% level allowed
by current LHC data.  
The only way to cancel this correction is to turn on the mixing, which
effectively suppresses the coupling of the light stop to the Higgs.  
But this is in conflict with the requirements of the EWB scenario,
which requires a near-maximal stop-Higgs coupling and therefore
small mixing.  
As we will see in Sec.~\ref{sec:results}, this tension cannot be reconciled with the current data.  

Before proceeding, let us note that the constraints from electroweak
precision fits and the LHC Higgs measurements largely overlap,
pointing to the same region in the stop parameter
space~\cite{Espinosa:2012in,Craig:2014una,Fan:2014axa}. In this region, called 
``funnel regime'' or ``blind spot'' in the literature, the shift in
the $hgg/h\gamma\gamma$ couplings and the stop contribution to the T
parameter are both 
minimized. (The stop contribution to the S parameter is small and plays a
subdominant role in the fits.) This occurs at approximately 
\beq
\sin (2\theta) \approx \frac{2 m_t}{\mone - \mtwo} \ \ {\rm or} \ \
X_t \approx  \mone + \mtwo~.
\eeq 
In the limit $\mone\ll\mtwo$, relevant for the stop-catalyzed EWB, these conditions simply mean that the light stop coupling to the Higgs vanishes, up to terms of order $(\mone/\mtwo)^2$ (see Eq.~\leqn{eq:hst1st1} below). 

\section{Electroweak Phase Transition}
\label{sec:EWPT}
The viability of EWB requires that the EWPT be sufficiently strong, that is, that the rate for baryon number
changing sphaleron  transitions inside the broken
phase, $\Gamma_\mathrm{sph}$, be slow enough to avoid washout of the baryon
asymmetry. The rate $\Gamma_\mathrm{sph}$ is proportional to 
$\mathrm{exp}(-E_\mathrm{sph}/T)$, where $E_\mathrm{sph}$ is the
sphaleron energy at temperature $T$. The transition proceeds
when $T$ is below the bubble nucleation temperature, $T_N$, which is
generally just below the critical temperature $T_c$. The larger the
magnitude of $E_\mathrm{sph}/T_N$, the more effective will be the
baryon asymmetry preservation in the broken phase. 

In the context of perturbation theory, the computation of
$\Gamma_\mathrm{sph}$ entails considerable conceptual and theoretical
challenges~\cite{Patel:2011th,Cohen:2011ap}. A particularly
vexing one is maintenance of gauge-invariance. 
Loops containing gauge sector degrees of freedom (gauge bosons, unphysical scalars,
Fadeev-Popov ghosts)
introduce gauge-dependence into the finite temperature
effective action, $S_\mathrm{eff}(T)$. Obtaining a gauge-invariant
estimate of $\Gamma_\mathrm{sph}$ at $T_c$ is
possible~\cite{Patel:2011th}, but doing so requires a level of care not
typically followed in previous literature.  

Here we adopt a strategy that can be appropriate for the MSSM and other
scenarios wherein gauge degrees of freedom play a subdominant role in
generating the barrier between the symmetric and broken
vacua. Specifically, we truncate the one-loop effective potential
$V_\mathrm{eff}(T)$ at second order in the electroweak gauge couplings
$g$ and $g^\prime$ while retaining while retaining terms to all orders in the top-quark Yukawa coupling, $y_t$.
Doing so eliminates the
gauge-dependence that first arises at $\mathcal{O}(g^3)$ and that
comes in tandem with the gauge-loop contribution to the barrier
between the symmetric and broken-phase vacua. At the same time, it
retains the gauge-invariant stop contributions to the barrier that
enter first at $O(y_t^3)$ and that intuition tells us should dominate
the phase transition dynamics.  

This intuition is based on the stop contribution to the daisy resummation term in $V_\mathrm{eff}(T)$:
\beq
\label{eq:daisy}
\Delta V_\mathrm{daisy}^{\tilde t} (T) = -\frac{2N_C T}{12\pi}
\sum_{i=1,2}\left[ M_{\tilde t_i}(h, T)^3 - M_{\tilde t_i}(h)^3\right],
\eeq
where $N_C$ is the number of colors,  \lq\lq $h$" generically denotes
the vacuum expectation values of the two neutral doublet Higgses,
$M_{\tilde t_i}(h)$ is the zero-temperature mass of stop eigenstate
$t_i$ and $M_{\tilde t_i}(h, T)$ is the corresponding finite
temperature mass. When the lighter eigenstate is essentially the right-handed
stop, one has~\cite{Carena:1996wj} 
\begin{eqnarray}
M_{\tilde t_1}(h, T)^2 & \supset & y_t^2 h_u^2
\left(1-\frac{X_t^2}{m_Q^2}\right) + m_U^2 +\Pi_{\tilde t_1}(T),
\end{eqnarray}
where $m_Q$ and $m_U$ are the left- and right-handed stop soft mass
parameters, respectively, 
and  
$\Pi_{\tilde t_1}(T)$ is the one-loop thermal contribution to the
stop mass-squared. Choosing $m_U^2\approx - \Pi_{\tilde t_1}(T)$
mitigates the screening of the stop contribution due to the daisy
resummation. The resulting  approximate $-T h_u^3$ term in the
potential increases the barrier between broken and unbroken vacua,
lowers $T_c$, and increases the ratio $E_\mathrm{sph}/T$ as needed for
baryon number preservation\cite{Carena:1996wj}. For $T\sim 100$ GeV,
this choice leads to a lightest stop mass on the order of 100
GeV. Note that the coefficient of the stop-induced $-T h_u^3$ term is
enhanced by $2 N_C y_t^3$. The gauge sector contributions, which are
not included due to our truncation, carry no such enhancement. 

The requirements for effective baryon number preservation follow from
solving the sphaleron equations of motion and computing
$E_\mathrm{sph}$. We observe that a consistent, non-trivial solution
of these equations requires retaining gauge contributions to {\em at
  least} $\mathcal{O}(g^2)$ since the Higgs quartic self-couplings
that enter the tree-level potential are $\mathcal{O}(g^2)$ as is the
coupling between the gauge field and Higgs profile
functions\cite{Moreno:1996zm}. In the present set-up, we formally
retain all $\mathcal{O}(g^2)$ contributions, but include none at
higher order in $g$ so as to maintain gauge invariance and consistency
of the sphaleron equations of motion. In practice, 
for simplicity of numerical analysis
we have not included the electroweak gauge boson contributions to the
thermal masses that are also second order in $g$. We have estimated
that doing so would result in shifts in the crucial quantity $\xi$, defined in Eq.~\eqref{eq:bnpc}, by no more than
10\%, leaving our conclusions unaffected. 

The resulting baryon number preservation criterion can be expressed as a condition on the ratio\cite{Quiros:1999jp}
\beq
\label{eq:bnpc}
\xi \equiv \frac{v(T_c)}{T_c}\gtrsim 1\ \ \ ,
\eeq
where $v(T_c)$ is the value of $h(T)$ that minimizes
$V_\mathrm{eff}(T)$ at the critical temperature. As discussed in
Ref.~\cite{Patel:2011th}, there exist numerous sources of uncertainty
in this condition, including the duration of the phase transition, the value of the
baryon asymmetry at the start of the transition, the computation of
the sphaleron fluctuation determinant, the origin of the unstable mode
of the sphaleron, and neglected higher order
loops~\cite{Espinosa:1996qw,Carena:1997ki,Cohen:2011ap}. Consequently, 
the precise numerical 
results should be taken with a grain of salt. Nevertheless, we believe that our
qualitative conclusions will not be altered, even taking into account
significant uncertainties associated with the use of perturbation
theory to analyze the phase transition dynamics. To that end, we will
show on our plots the contours of $\xi = 0.5$ and $\xi = 0$ (corresponding to the absence of a first order transition) to illustrate the potential impact of these uncertainties.

\begin{figure}[t!]
\centering
\includegraphics[width=.42\textwidth]{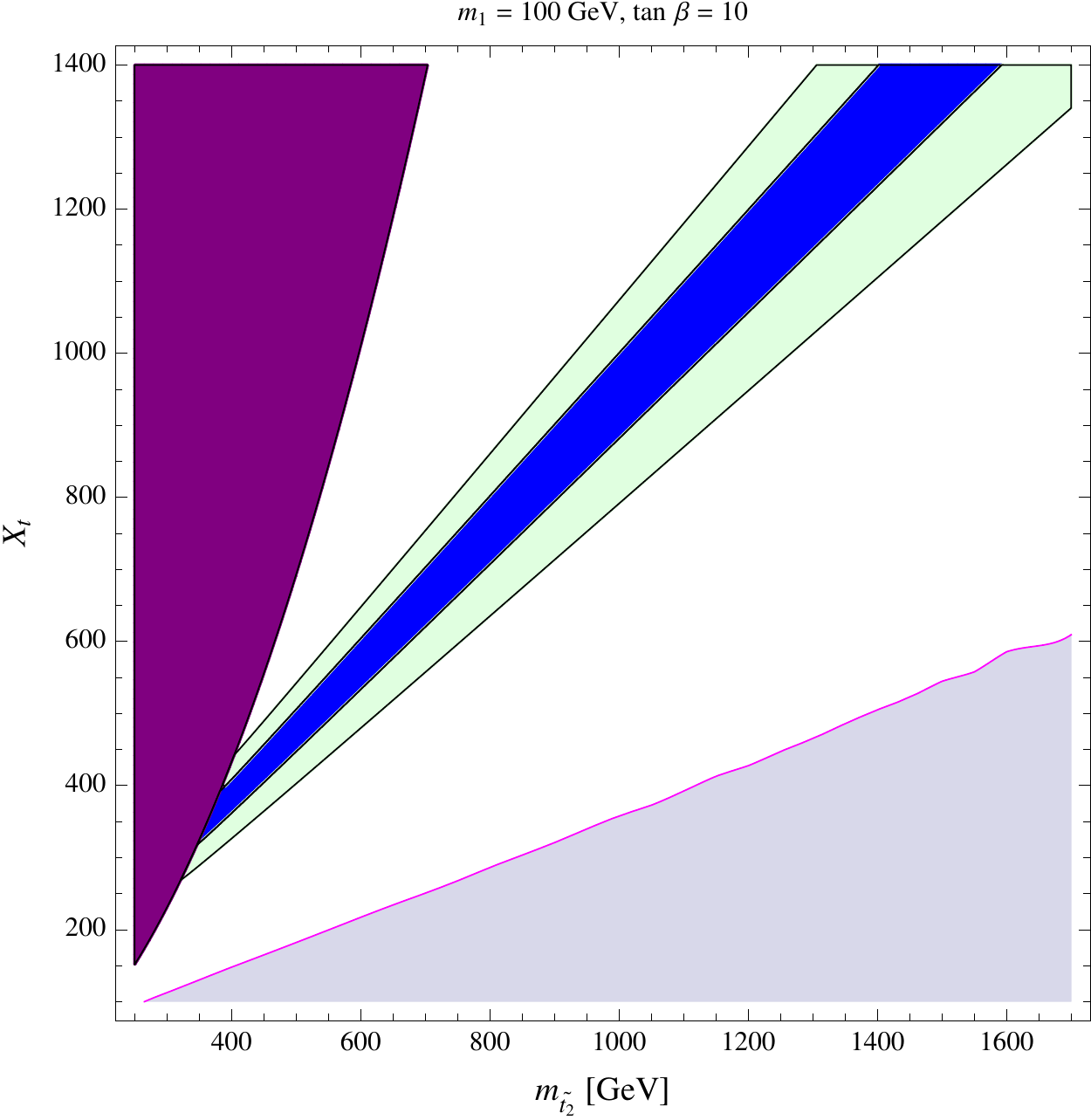}
\caption{Regions of stop parameter space allowed by the LHC Higgs
  measurements (blue - 67\%~CL and green - 95\%~CL) vs. the domain
  where the stop-catalyzed EWB can potentially be viable ($\xi > 0$)
in pink. Mass of the light stop is fixed at 100~GeV. The unphysical
region (no solution for $\theta_t$) is shaded
in purple. } 
\label{fig:100xt}
\end{figure}

\section{Results}
\label{sec:results}
We performed a numerical scan over the four-dimensional parameter
space outlined in sec.~\ref{sec:SUSY}. For each point in the scan, we
evaluated consistency with the experimental constraints by performing
a fit to the LHC Higgs measurements, using the data set and the
fitting procedure of Ref.~\cite{Giardino:2013bma}, and the expressions
for the coupling shifts from Ref.~\cite{Espinosa:2012in}. (For details of the fit, see Appendix A.) We also
evaluated the constraints from EW precision measurements; however, we
find these to be consistently weaker that the Higgs fit
constraints. 
Furthermore, for each point in the scan, we determined whether or not
the EWPT is strongly 1st-order, using the procedure outlined in the previous section. 

We find that the 1st-order phase transition requires a very light
stop, $\mone\simlt 110$ GeV, independent of the other parameters. On
the other hand, LEP-2 constraints imply $\mone\simgt100$ GeV,
confining this parameter to a narrow band. Within this band, the Higgs
fit constraints on the remaining parameters vary only slightly with
$\mone$. In the plots below, we choose $\mone=100$ GeV as a
representative value, but the picture that emerges from these plots is
valid throughout the allowed range of $\mone$. Likewise, we fix
$\tan\beta=10$ in the plots as a representative value. For larger
$\tan\beta$ values our  
results stay almost independent of $\tan\beta$, while for lower
$\tan \beta$ the EWPT becomes weaker while the Higgs constraints are
largely unaffected. Again, the picture that emerges 
remains valid independent of $\tan\beta$.       

\begin{figure}[h]
\centering
\includegraphics[width=.42\textwidth]{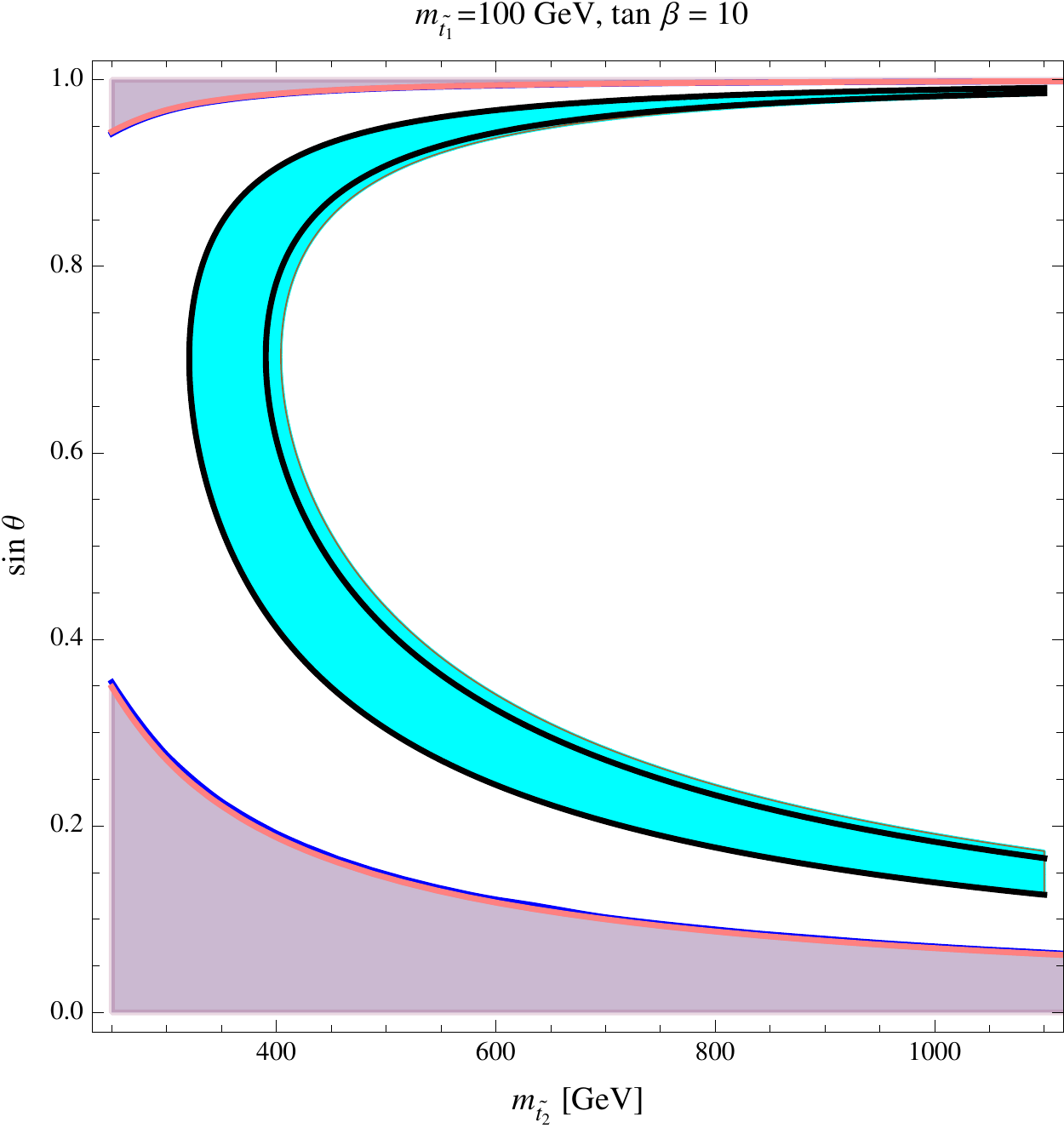}
\caption{Constraints and regions with a 1st-order EWPT in the ($\mtwo,
  \theta_t$) plane, for the light stop mass $\mone=100$ GeV.  The
  region allowed by Higgs fits at 95\% CL is shaded in
  blue. The blue 
  line shows the contour of $\xi = 0$ and the red line shows the
  contour of $\xi = 0.5$. The region between the black contours is allowed at 
  the 95\% CL if a non-zero Higgs invisible width is included ($\epsilon_{inv} = 0.1$
  using the definitions of~\cite{Giardino:2013bma}).}
\label{fig:100stop}
\end{figure}

\begin{figure}[h]
\centering
\includegraphics[width=.45\textwidth]{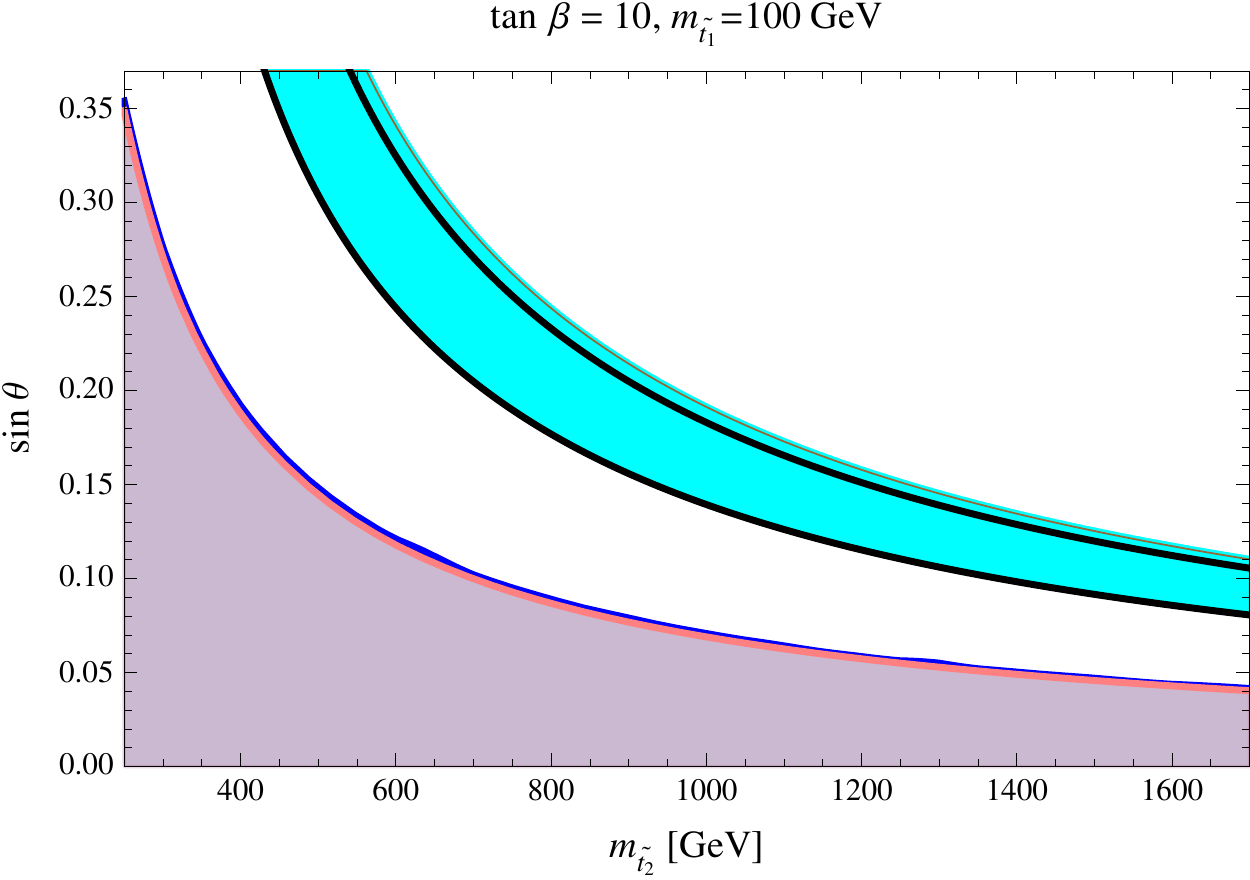}
\includegraphics[width=.45\textwidth]{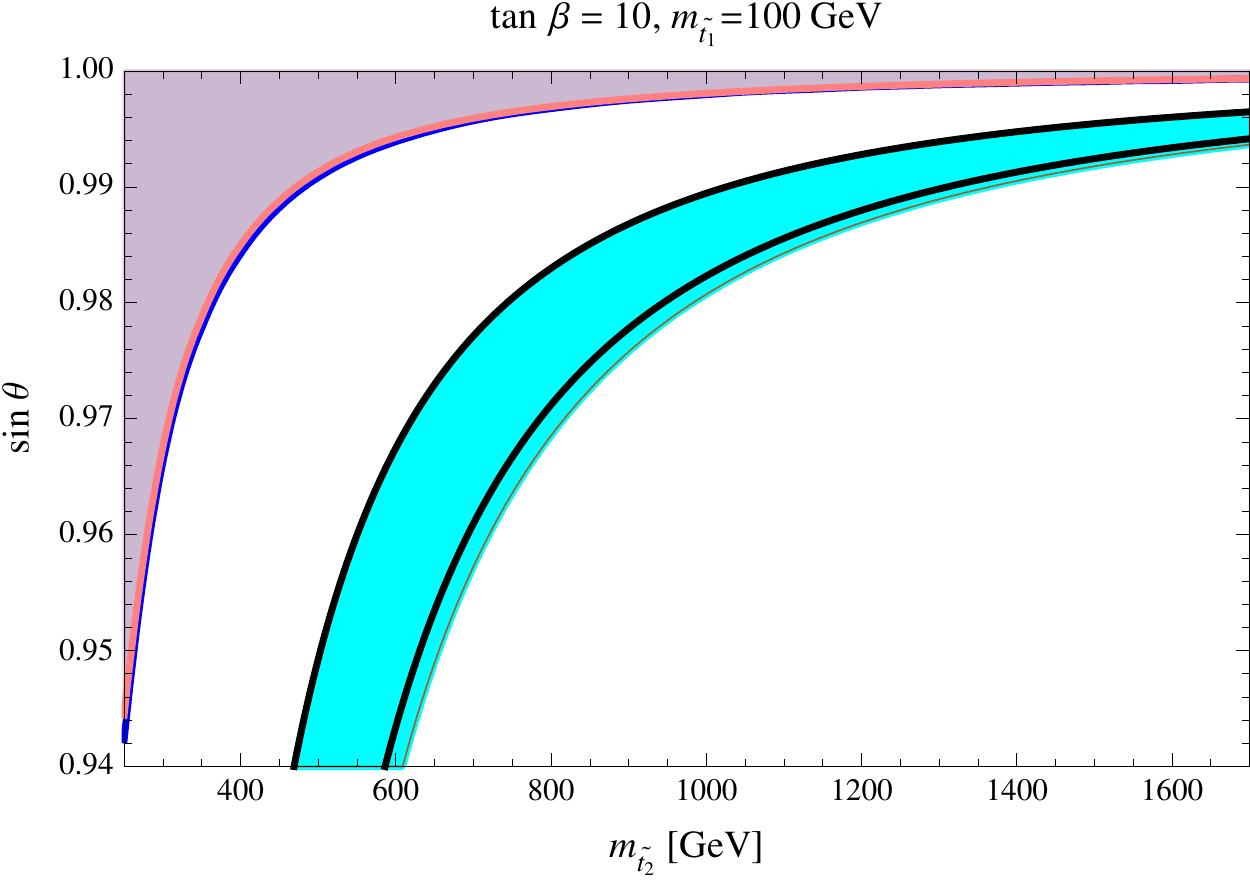}
\caption{Same as Fig.~\ref{fig:100stop}, zooming in on the
  regions where the light 
stop is mostly $\tilde{t}_L$ and $\tilde{t}_R$.}
\label{fig:RHstop100}
\end{figure}

The main results of our analysis are summarized in
Figs.~\ref{fig:100xt},~\ref{fig:100stop} and~\ref{fig:RHstop100}. The conclusion is
clear: {\it there is no overlap between the parameter space regions
  allowed by the Higgs fit, and those consistent with a 1st-order
  EWPT within a perturbative calculation.} Thus, the stop-catalyzed
EWB scenario is no longer 
viable. Perhaps the clearest way to understand this result is provided
by Fig.~\ref{fig:100xt}. Not surprisingly, the region of parameter
space allowed by Higgs measurements is a band around the line
$X_t\approx m_{\tilde t_2}$: this follows directly from
Eq.~\eqref{eq:hgg} in the limit $\mone\ll\mtwo$. The crucial
observation is that \emph{along the contours 
  of constant EWPT strength $\xi$, $X_t$ also scales linearly with $m_{\tilde
    t_2}$}. This is related to the fact that the effective coupling
between the lightest stop and the Higgs is given by 
\beq\label{eq:hst1st1} 
\cL_{eff} =  y_t^2 \left( 1- \frac{X_t^2}{\mtwo^2-\mone^2} \right) |H|^2
|\tilde t_1|.
\eeq 
The thermal potential is determined almost exclusively by this
effective coupling, so that constant-$\xi$ contours in the regime
$\mone\ll\mtwo$ correspond to a fixed ratio $X_t/\mtwo$. Crucially, a 
1st-order transition is only possible when the effective coupling is
close to 1; specifically, $X_t/\mtwo \simlt 0.3 $ is required, as can
be seen in Fig.~\ref{fig:100xt}. This region does not overlap with the
region $X_t/\mtwo\approx 1$ allowed by the Higgs fits, regardless of
the value of $\mtwo$. Incidentally, this argument provides a clear
understanding of the results of Ref.~\cite{Curtin:2012aa} regarding
the MSSM, where $\mtwo\sim100$ TeV is required by the 125 GeV Higgs
mass.   

Another useful representation of the same results is shown in
Figs.~\ref{fig:100stop} and~\ref{fig:RHstop100}, where the $X_t$ parameter has been traded
for the stop mixing angle $\theta_t$. These plots make it clear for a
100 GeV light stop, the Higgs fits imply a tight relationship between
$\theta_t$ and $\mtwo$, which unfortunately is incompatible with the
small-mixing regime required by the stop-catalyzed EWB.

Now we address an important potential caveat to the above
argument. The Higgs coupling fit presented above is performed under
the assumption of no new Higgs decay channels beyond those present in
the SM. Lifting this assumption may be expected to relax the Higgs fit
constraints. In particular, the strong constraint on the gluon
coupling comes from the agreement of the inferred production cross
section in the dominant $gg\to h$ channel with the SM. If the Higgs
has a new decay channel, a production cross section   
in excess of the SM value can be accommodated without changing the
observed event rates in any of the channels. In the stop-catalyzed EWB
scenario, it is quite natural for the Higgs to have an appreciable
decay width into light neutralino pairs, resulting in an invisible
final state. It was argued in Ref.~\cite{Carena:2012np}  
that this may revive the stop-catalyzed EWB scenario, even within the
MSSM.  

However, since Ref.~\cite{Carena:2012np} appeared, dedicated
experimental searches for Higgs invisible decays have been 
performed by both ATLAS and CMS, which strongly constrain this
possibility~\cite{ATLAS-CONF-2013-011,CMS-PAS-HIG-13-018}. 
Ref.~\cite{Giardino:2013bma} combined two these measurements (both of
which had downward fluctuations) 
and estimated the allowed Higgs invisible branching ratio as $-0.18\pm0.31$, of 
course perfectly consistent with zero. The bound on $h \gamma \gamma $ coupling has also been improved significantly, so that the gluon coupling does not dominate the fit as strongly as in the early analyses of the LHC Higgs data. Due to both these
factors, turning on the Higgs invisible width no longer improves the overall Higgs couplings fit, and does not
relax the constraints in the ``right'' direction for the stop-catalyzed EWB scenario. We explicitly illustrate this point
in Figs.~\ref{fig:100stop} and~\ref{fig:RHstop100}, which shows that with the current data, allowing for the Higgs invisible width does not improve
the fit.



\section{Conclusions}
\label{sec:conc}

While stop-catalyzed EWB is a theoretically attractive scenario, it
has been known for some time that the LHC measurements of the Higgs
mass and rates rule it out in the case of the MSSM. In this paper, we
extended the analysis to supersymmetric models with non-minimal
contributions to the Higgs potential. Such contributions    
can lift the tree-level Higgs mass above the MSSM bound, thus
eliminating one of the most stringent MSSM constraints on the stop
sector parameters. One might hope that the newly expanded parameter
space would include regions compatible with a stop-catalyzed EWB. We
showed that this is unfortunately not the case. The reason for this is
simple: the stop-catalyzed EWB requires a light stop ($\mone\sim 100$
GeV) with near-maximal coupling to the Higgs, since otherwise the
effects of stop loops on the Higgs thermal potential are not large
enough to trigger a strong 1st-order electroweak phase transition. On the
other hand, such a light stop is compatible with the LHC constraints
on the $hgg$ and $h\gamma\gamma$ couplings only if its coupling to the
Higgs is suppressed, and is far from maximal. We quantified these
requirements and found no overlap between the parameter space regions
with viable stop-catalyzed EWB and those with Higgs couplings
compatible with the LHC data. The conclusion holds even in the
presence of non-SM invisible Higgs decay channels such as $h\to
\tilde{\chi}^0 \tilde{\chi}^0$.  

An important limitation of our analysis is the assumption that no
additional scalars beyond a single SM-like Higgs participate in the
electroweak phase transition. This assumption can be violated in
non-minimal supersymmetric models  with extended Higgs sectors. For
example, in the NMSSM-like scenario, an additional gauge-singlet
scalar field can be 
active during the phase transition, leading to novel possibilities
such as a two-step phase transition. It has been shown that a strong 
1st-order electroweak phase transition remains a viable possibility in
this class of models~\cite{Pietroni:1992in,Menon:2004wv,Huber:2006wf}. In these
cases, the potential barrier necessary for the strong 1st-order transition
typically arises from the tree-level multi-field scalar potential,
rather than thermal loop effects, making them quite distinct from the
stop-catalyzed scenario we focused on here. Still, these models
demonstrate that even though the stop-catalyzed EWB no longer appears
viable, other options for successful EWB exist in the supersymmetric
context.    

The analysis of this paper provides an illustration of the power of
the Higgs data collected by the LHC to shed light on the electroweak
phase transition, an important event in the history of the Universe
about which we currently have no direct information. The current
10-20\% accuracy of the Higgs coupling measurements is already
sufficient to rule out one of the most popular scenarios with a strong
1st-order phase transition. Many other models (both SUSY and non-SUSY)
that allow for a strong 1st-order phase transition predict more subtle
deviations from the SM, and testing these models will require
increasing the precision of the Higgs coupling measurements to the 1\%
level and beyond, as well as measuring the Higgs cubic
self-coupling~\cite{Noble:2007kk,Katz:2014bha,Profumo:2014opa,Curtin:2014jma}. The  
proposed electron-positron Higgs factories and a 100 TeV proton-proton
collider will be needed to comprehensively probe the viability of a
strong 
1st-order EWPT, and hence of electroweak baryogenesis.

\acknowledgments{We are grateful to Roberto Franceschini, Michelangelo
  Mangano and Veronica
  Sanz   for 
  useful discussions. The work of AK was accomplished at the
  Aspen Center for Physics, which is supported by National Science
  Foundation grant PHY-1066293. The research of AK and MJRM was also
  supported by the 
  Munich Institute for Astro- and Particle Physics (MIAPP) of the DFG
  cluster of excellence "Origin and Structure of the Universe". MP is
  supported by the U.S. National Science Foundation through grant
  PHY-1316222. MJRM and PW were supported in part by U.S. Department
  of Energy contract DE-SC0011095.}  

\appendix 

\section{Details of the Higgs Couplings Fit}

Constraints on the stop sector parameters were derived from a fit to the Higgs couplings measured at the LHC. In this Appendix we describe the fitting procedure in more detail. Since the procedure closely follows the prescription of
Ref.~\cite{Giardino:2013bma}, one can view this Appendix as a short
executive summary of this reference. 

The Higgs couplings taken into account in the fit are: $hb\bar
b$, $hgg$, $h\to {\rm inv}$, $hWW$, $hZZ$, $h \gamma \gamma$ and $h
\tau \tau$. In the main part of our analysis, which assumes that the 2HDM is in the decoupling regime, the only deviations are in $h
\gamma \gamma$ and $hgg$ and, in the case when Higgs decay to a
neutralino pair is allowed, in $h\to {\rm inv}$. The rest of the couplings only 
become important when we explicitly check the effects of 2HDM outside of the decoupling regime, which neccessarily
triggers deviations in the couplings to the down type sector fermions
and to the lesser extend to the gauge bosons. 

First, we slightly simplify the ``proper'' fit, by assuming
that all error bars in the LHC measurements are Gaussian (which
does not lead to a significant loss of information). Second, we assume that all
deviations from the SM-predicted values are small, namely $r_i = 1
+ \epsilon_i$, where $r_i$'s are defined as 
\beq
\cL_{higgs} & = & r_\psi \frac{m_\psi}{v} h \bar \psi \psi + r_Z
\frac{M_Z^2}{v} h Z^\mu Z_\mu +   
 r_W \frac{2 m_W^2}{v} h W_\mu W^\mu + \nonumber \\ && r_\gamma c_{SM}^{\gamma
  \gamma} \frac{\alpha}{\pi v} h F_{\mu \nu}^2 + r_g c_{SM}^{gg}
 \frac{\alpha_s}{\pi v} h G_{\mu \nu}^2~,  
\eeq  
and $\epsilon_i \ll 1$. For the invisible rate, which is expected to
vanish in the SM, we assume ${\rm BR} (h \to {\rm inv}) =
\epsilon_{inv} $. To estimate the measured deviation of the $\gamma
\gamma$ coupling from the SM we use the results of
ATLAS~\cite{ATLAS-CONF-2013-012} and 
CMS~\cite{Khachatryan:2014ira}. By combining these results we estimate
$\epsilon_\gamma = 0.00 \pm 0.14$ (of course we quote
  $1\sigma$ uncertainties). The coupling $hgg$ is extracted from all the exclusive (gluon fusion)
decay modes, namely $\gamma
\gamma$~\cite{ATLAS-CONF-2013-012,Khachatryan:2014ira},
$ZZ^*$~\cite{Aad:2013wqa,Chatrchyan:2013mxa},
$WW^*$~\cite{ATLAS-CONF-2013-030, Chatrchyan:2013iaa}, and to the
lesser extent from the fermionic channels. The average is $\epsilon_g
= -0.13 \pm 0.20$. For the extraction of the bounds on the deviations of the other couplings, which are
less important in our fit, the reader is again referred to
Ref.~\cite{Giardino:2013bma}. In the basis $\epsilon_i = (\epsilon_b, \epsilon_g,
\epsilon_{inv}, \epsilon_W, \epsilon_Z, \epsilon_\gamma,
\epsilon_\tau)$, the correlation matrix between the
different $\epsilon_i$'s, based on the theoretical calculation, is  
\beq
\rho = \left(
\begin{array}{ccccccc}
1 & 0.70 &  0.04 &  0.52 & 0.38 & 0.58 & 0.59 \\
0.70 & 1 &  0.43 & 0.38 & 0.11 & 0.40 & 0.52 \\
0.04 & 0.43 & 1 & 0.46 & 0.13 & 0.40 & 0.34 \\
0.52 &  0.38 &  0.46 & 1 & 0.44 & 0.63 & 0.45 \\
0.38 &  0.11 &  0.13 & 0.44 & 1 &  0.42 & 0.33 \\ 
0.58  & 0.40 &  0.40 &  0.63 &  0.42 &  1  & 0.54 \\ 
0.59 & 0.52 & 0.34 & 0.45 & 0.33 & 0.54 & 1 \\
\end{array} 
\right)
\eeq
For completeness we also quote here the allowed ranges
of the other $\epsilon$'s, as calculated by~\cite{Giardino:2013bma}:
\beq
\epsilon_b & = & -0.19 \pm 0.28~, \\
\epsilon_{inv} & = & -0.22 \pm 0.20~, \\
\epsilon_{W} & = & -0.2 \pm 0.13~, \\
\epsilon_Z & = & 0.00 \pm 0.10~, \\
\epsilon_\tau & = & -0.03 \pm 0.17~. 
\eeq
Again, these channels were important in the fit only when we considered the non-decoupling limit of the 2HDM and/or
the possibility of an invisible decay of the Higgs to light neutralinos.

\bibliography{refs}

\begin{thebibliography}{72}
\expandafter\ifx\csname natexlab\endcsname\relax\def\natexlab#1{#1}\fi
\expandafter\ifx\csname bibnamefont\endcsname\relax
  \def\bibnamefont#1{#1}\fi
\expandafter\ifx\csname bibfnamefont\endcsname\relax
  \def\bibfnamefont#1{#1}\fi
\expandafter\ifx\csname citenamefont\endcsname\relax
  \def\citenamefont#1{#1}\fi
\expandafter\ifx\csname url\endcsname\relax
  \def\url#1{\texttt{#1}}\fi
\expandafter\ifx\csname urlprefix\endcsname\relax\def\urlprefix{URL }\fi
\providecommand{\bibinfo}[2]{#2}
\providecommand{\eprint}[2][]{\url{#2}}

\bibitem[{\citenamefont{Morrissey and Ramsey-Musolf}(2012)}]{Morrissey:2012db}
\bibinfo{author}{\bibfnamefont{D.~E.} \bibnamefont{Morrissey}}
  \bibnamefont{and} \bibinfo{author}{\bibfnamefont{M.~J.}
  \bibnamefont{Ramsey-Musolf}}, \emph{\bibinfo{title}{{Electroweak
  baryogenesis}}}, \bibinfo{journal}{New J.Phys.}
  \textbf{\bibinfo{volume}{14}}, \bibinfo{pages}{125003}
  (\bibinfo{year}{2012}), \eprint{1206.2942}.

\bibitem[{\citenamefont{Sakharov}(1967)}]{Sakharov:1967dj}
\bibinfo{author}{\bibfnamefont{A.}~\bibnamefont{Sakharov}},
  \emph{\bibinfo{title}{{Violation of CP Invariance, c Asymmetry, and Baryon
  Asymmetry of the Universe}}}, \bibinfo{journal}{Pisma Zh.Eksp.Teor.Fiz.}
  \textbf{\bibinfo{volume}{5}}, \bibinfo{pages}{32} (\bibinfo{year}{1967}).

\bibitem[{\citenamefont{Gurtler et~al.}(1997)\citenamefont{Gurtler, Ilgenfritz,
  and Schiller}}]{Gurtler:1997hr}
\bibinfo{author}{\bibfnamefont{M.}~\bibnamefont{Gurtler}},
  \bibinfo{author}{\bibfnamefont{E.-M.} \bibnamefont{Ilgenfritz}},
  \bibnamefont{and} \bibinfo{author}{\bibfnamefont{A.}~\bibnamefont{Schiller}},
  \emph{\bibinfo{title}{{Where the electroweak phase transition ends}}},
  \bibinfo{journal}{Phys. Rev.} \textbf{\bibinfo{volume}{D56}},
  \bibinfo{pages}{3888} (\bibinfo{year}{1997}), \eprint{hep-lat/9704013}.

\bibitem[{\citenamefont{Laine and Rummukainen}(1999)}]{Laine:1998jb}
\bibinfo{author}{\bibfnamefont{M.}~\bibnamefont{Laine}} \bibnamefont{and}
  \bibinfo{author}{\bibfnamefont{K.}~\bibnamefont{Rummukainen}},
  \emph{\bibinfo{title}{{What's new with the electroweak phase transition?}}},
  \bibinfo{journal}{Nucl. Phys. Proc. Suppl.} \textbf{\bibinfo{volume}{73}},
  \bibinfo{pages}{180} (\bibinfo{year}{1999}), \eprint{hep-lat/9809045}.

\bibitem[{\citenamefont{Csikor et~al.}(1999)\citenamefont{Csikor, Fodor, and
  Heitger}}]{Csikor:1998eu}
\bibinfo{author}{\bibfnamefont{F.}~\bibnamefont{Csikor}},
  \bibinfo{author}{\bibfnamefont{Z.}~\bibnamefont{Fodor}}, \bibnamefont{and}
  \bibinfo{author}{\bibfnamefont{J.}~\bibnamefont{Heitger}},
  \emph{\bibinfo{title}{{Endpoint of the hot electroweak phase transition}}},
  \bibinfo{journal}{Phys. Rev. Lett.} \textbf{\bibinfo{volume}{82}},
  \bibinfo{pages}{21} (\bibinfo{year}{1999}), \eprint{hep-ph/9809291}.

\bibitem[{\citenamefont{Aoki et~al.}(1999)\citenamefont{Aoki, Csikor, Fodor,
  and Ukawa}}]{Aoki:1999fi}
\bibinfo{author}{\bibfnamefont{Y.}~\bibnamefont{Aoki}},
  \bibinfo{author}{\bibfnamefont{F.}~\bibnamefont{Csikor}},
  \bibinfo{author}{\bibfnamefont{Z.}~\bibnamefont{Fodor}}, \bibnamefont{and}
  \bibinfo{author}{\bibfnamefont{A.}~\bibnamefont{Ukawa}},
  \emph{\bibinfo{title}{{The Endpoint of the first order phase transition of
  the SU(2) gauge Higgs model on a four-dimensional isotropic lattice}}},
  \bibinfo{journal}{Phys. Rev.} \textbf{\bibinfo{volume}{D60}},
  \bibinfo{pages}{013001} (\bibinfo{year}{1999}), \eprint{hep-lat/9901021}.

\bibitem[{\citenamefont{Gavela et~al.}(1994{\natexlab{a}})\citenamefont{Gavela,
  Hernandez, Orloff, and Pene}}]{Gavela:1993ts}
\bibinfo{author}{\bibfnamefont{M.~B.} \bibnamefont{Gavela}},
  \bibinfo{author}{\bibfnamefont{P.}~\bibnamefont{Hernandez}},
  \bibinfo{author}{\bibfnamefont{J.}~\bibnamefont{Orloff}}, \bibnamefont{and}
  \bibinfo{author}{\bibfnamefont{O.}~\bibnamefont{Pene}},
  \emph{\bibinfo{title}{{Standard model CP violation and baryon asymmetry}}},
  \bibinfo{journal}{Mod. Phys. Lett.} \textbf{\bibinfo{volume}{A9}},
  \bibinfo{pages}{795} (\bibinfo{year}{1994}{\natexlab{a}}),
  \eprint{hep-ph/9312215}.

\bibitem[{\citenamefont{Huet and Sather}(1995)}]{Huet:1994jb}
\bibinfo{author}{\bibfnamefont{P.}~\bibnamefont{Huet}} \bibnamefont{and}
  \bibinfo{author}{\bibfnamefont{E.}~\bibnamefont{Sather}},
  \emph{\bibinfo{title}{{Electroweak baryogenesis and standard model CP
  violation}}}, \bibinfo{journal}{Phys. Rev.} \textbf{\bibinfo{volume}{D51}},
  \bibinfo{pages}{379} (\bibinfo{year}{1995}), \eprint{hep-ph/9404302}.

\bibitem[{\citenamefont{Gavela et~al.}(1994{\natexlab{b}})\citenamefont{Gavela,
  Hernandez, Orloff, Pene, and Quimbay}}]{Gavela:1994dt}
\bibinfo{author}{\bibfnamefont{M.~B.} \bibnamefont{Gavela}},
  \bibinfo{author}{\bibfnamefont{P.}~\bibnamefont{Hernandez}},
  \bibinfo{author}{\bibfnamefont{J.}~\bibnamefont{Orloff}},
  \bibinfo{author}{\bibfnamefont{O.}~\bibnamefont{Pene}}, \bibnamefont{and}
  \bibinfo{author}{\bibfnamefont{C.}~\bibnamefont{Quimbay}},
  \emph{\bibinfo{title}{{Standard model CP violation and baryon asymmetry. Part
  2: Finite temperature}}}, \bibinfo{journal}{Nucl. Phys.}
  \textbf{\bibinfo{volume}{B430}}, \bibinfo{pages}{382}
  (\bibinfo{year}{1994}{\natexlab{b}}), \eprint{hep-ph/9406289}.

\bibitem[{\citenamefont{Li et~al.}(2009)\citenamefont{Li, Profumo, and
  Ramsey-Musolf}}]{Li:2008ez}
\bibinfo{author}{\bibfnamefont{Y.}~\bibnamefont{Li}},
  \bibinfo{author}{\bibfnamefont{S.}~\bibnamefont{Profumo}}, \bibnamefont{and}
  \bibinfo{author}{\bibfnamefont{M.}~\bibnamefont{Ramsey-Musolf}},
  \emph{\bibinfo{title}{{Bino-driven Electroweak Baryogenesis with highly
  suppressed Electric Dipole Moments}}}, \bibinfo{journal}{Phys. Lett.}
  \textbf{\bibinfo{volume}{B673}}, \bibinfo{pages}{95} (\bibinfo{year}{2009}),
  \eprint{0811.1987}.

\bibitem[{\citenamefont{Cirigliano et~al.}(2010)\citenamefont{Cirigliano, Li,
  Profumo, and Ramsey-Musolf}}]{Cirigliano:2009yd}
\bibinfo{author}{\bibfnamefont{V.}~\bibnamefont{Cirigliano}},
  \bibinfo{author}{\bibfnamefont{Y.}~\bibnamefont{Li}},
  \bibinfo{author}{\bibfnamefont{S.}~\bibnamefont{Profumo}}, \bibnamefont{and}
  \bibinfo{author}{\bibfnamefont{M.~J.} \bibnamefont{Ramsey-Musolf}},
  \emph{\bibinfo{title}{{MSSM Baryogenesis and Electric Dipole Moments: An
  Update on the Phenomenology}}}, \bibinfo{journal}{JHEP}
  \textbf{\bibinfo{volume}{01}}, \bibinfo{pages}{002} (\bibinfo{year}{2010}),
  \eprint{0910.4589}.

\bibitem[{\citenamefont{Kozaczuk et~al.}(2012)\citenamefont{Kozaczuk, Profumo,
  Ramsey-Musolf, and Wainwright}}]{Kozaczuk:2012xv}
\bibinfo{author}{\bibfnamefont{J.}~\bibnamefont{Kozaczuk}},
  \bibinfo{author}{\bibfnamefont{S.}~\bibnamefont{Profumo}},
  \bibinfo{author}{\bibfnamefont{M.~J.} \bibnamefont{Ramsey-Musolf}},
  \bibnamefont{and} \bibinfo{author}{\bibfnamefont{C.~L.}
  \bibnamefont{Wainwright}}, \emph{\bibinfo{title}{{Supersymmetric Electroweak
  Baryogenesis Via Resonant Sfermion Sources}}}, \bibinfo{journal}{Phys. Rev.}
  \textbf{\bibinfo{volume}{D86}}, \bibinfo{pages}{096001}
  (\bibinfo{year}{2012}), \eprint{1206.4100}.

\bibitem[{\citenamefont{Carena et~al.}(1996)\citenamefont{Carena, Quiros, and
  Wagner}}]{Carena:1996wj}
\bibinfo{author}{\bibfnamefont{M.}~\bibnamefont{Carena}},
  \bibinfo{author}{\bibfnamefont{M.}~\bibnamefont{Quiros}}, \bibnamefont{and}
  \bibinfo{author}{\bibfnamefont{C.~E.~M.} \bibnamefont{Wagner}},
  \emph{\bibinfo{title}{{Opening the window for electroweak baryogenesis}}},
  \bibinfo{journal}{Phys. Lett.} \textbf{\bibinfo{volume}{B380}},
  \bibinfo{pages}{81} (\bibinfo{year}{1996}), \eprint{hep-ph/9603420}.

\bibitem[{\citenamefont{Delepine et~al.}(1996)\citenamefont{Delepine, Gerard,
  Gonzalez~Felipe, and Weyers}}]{Delepine:1996vn}
\bibinfo{author}{\bibfnamefont{D.}~\bibnamefont{Delepine}},
  \bibinfo{author}{\bibfnamefont{J.~M.} \bibnamefont{Gerard}},
  \bibinfo{author}{\bibfnamefont{R.}~\bibnamefont{Gonzalez~Felipe}},
  \bibnamefont{and} \bibinfo{author}{\bibfnamefont{J.}~\bibnamefont{Weyers}},
  \emph{\bibinfo{title}{{A Light stop and electroweak baryogenesis}}},
  \bibinfo{journal}{Phys. Lett.} \textbf{\bibinfo{volume}{B386}},
  \bibinfo{pages}{183} (\bibinfo{year}{1996}), \eprint{hep-ph/9604440}.

\bibitem[{\citenamefont{Carena et~al.}(2009)\citenamefont{Carena, Nardini,
  Quiros, and Wagner}}]{Carena:2008vj}
\bibinfo{author}{\bibfnamefont{M.}~\bibnamefont{Carena}},
  \bibinfo{author}{\bibfnamefont{G.}~\bibnamefont{Nardini}},
  \bibinfo{author}{\bibfnamefont{M.}~\bibnamefont{Quiros}}, \bibnamefont{and}
  \bibinfo{author}{\bibfnamefont{C.}~\bibnamefont{Wagner}},
  \emph{\bibinfo{title}{{The Baryogenesis Window in the MSSM}}},
  \bibinfo{journal}{Nucl.Phys.} \textbf{\bibinfo{volume}{B812}},
  \bibinfo{pages}{243} (\bibinfo{year}{2009}), \eprint{0809.3760}.

\bibitem[{\citenamefont{Pietroni}(1993)}]{Pietroni:1992in}
\bibinfo{author}{\bibfnamefont{M.}~\bibnamefont{Pietroni}},
  \emph{\bibinfo{title}{{The Electroweak phase transition in a nonminimal
  supersymmetric model}}}, \bibinfo{journal}{Nucl. Phys.}
  \textbf{\bibinfo{volume}{B402}}, \bibinfo{pages}{27} (\bibinfo{year}{1993}),
  \eprint{hep-ph/9207227}.

\bibitem[{\citenamefont{Menon et~al.}(2004)\citenamefont{Menon, Morrissey, and
  Wagner}}]{Menon:2004wv}
\bibinfo{author}{\bibfnamefont{A.}~\bibnamefont{Menon}},
  \bibinfo{author}{\bibfnamefont{D.~E.} \bibnamefont{Morrissey}},
  \bibnamefont{and} \bibinfo{author}{\bibfnamefont{C.~E.~M.}
  \bibnamefont{Wagner}}, \emph{\bibinfo{title}{{Electroweak baryogenesis and
  dark matter in the nMSSM}}}, \bibinfo{journal}{Phys. Rev.}
  \textbf{\bibinfo{volume}{D70}}, \bibinfo{pages}{035005}
  (\bibinfo{year}{2004}), \eprint{hep-ph/0404184}.

\bibitem[{\citenamefont{Huber et~al.}(2006)\citenamefont{Huber, Konstandin,
  Prokopec, and Schmidt}}]{Huber:2006wf}
\bibinfo{author}{\bibfnamefont{S.~J.} \bibnamefont{Huber}},
  \bibinfo{author}{\bibfnamefont{T.}~\bibnamefont{Konstandin}},
  \bibinfo{author}{\bibfnamefont{T.}~\bibnamefont{Prokopec}}, \bibnamefont{and}
  \bibinfo{author}{\bibfnamefont{M.~G.} \bibnamefont{Schmidt}},
  \emph{\bibinfo{title}{{Electroweak Phase Transition and Baryogenesis in the
  nMSSM}}}, \bibinfo{journal}{Nucl. Phys.} \textbf{\bibinfo{volume}{B757}},
  \bibinfo{pages}{172} (\bibinfo{year}{2006}), \eprint{hep-ph/0606298}.

\bibitem[{\citenamefont{Engel et~al.}(2013)\citenamefont{Engel, Ramsey-Musolf,
  and van Kolck}}]{Engel:2013lsa}
\bibinfo{author}{\bibfnamefont{J.}~\bibnamefont{Engel}},
  \bibinfo{author}{\bibfnamefont{M.~J.} \bibnamefont{Ramsey-Musolf}},
  \bibnamefont{and} \bibinfo{author}{\bibfnamefont{U.}~\bibnamefont{van
  Kolck}}, \emph{\bibinfo{title}{{Electric Dipole Moments of Nucleons, Nuclei,
  and Atoms: The Standard Model and Beyond}}}, \bibinfo{journal}{Prog. Part.
  Nucl. Phys.} \textbf{\bibinfo{volume}{71}}, \bibinfo{pages}{21}
  (\bibinfo{year}{2013}), \eprint{1303.2371}.

\bibitem[{\citenamefont{Curtin et~al.}(2012)\citenamefont{Curtin, Jaiswal, and
  Meade}}]{Curtin:2012aa}
\bibinfo{author}{\bibfnamefont{D.}~\bibnamefont{Curtin}},
  \bibinfo{author}{\bibfnamefont{P.}~\bibnamefont{Jaiswal}}, \bibnamefont{and}
  \bibinfo{author}{\bibfnamefont{P.}~\bibnamefont{Meade}},
  \emph{\bibinfo{title}{{Excluding Electroweak Baryogenesis in the MSSM}}},
  \bibinfo{journal}{JHEP} \textbf{\bibinfo{volume}{1208}}, \bibinfo{pages}{005}
  (\bibinfo{year}{2012}), \eprint{1203.2932}.

\bibitem[{\citenamefont{Cohen et~al.}(2012)\citenamefont{Cohen, Morrissey, and
  Pierce}}]{Cohen:2012zza}
\bibinfo{author}{\bibfnamefont{T.}~\bibnamefont{Cohen}},
  \bibinfo{author}{\bibfnamefont{D.~E.} \bibnamefont{Morrissey}},
  \bibnamefont{and} \bibinfo{author}{\bibfnamefont{A.}~\bibnamefont{Pierce}},
  \emph{\bibinfo{title}{{Electroweak Baryogenesis and Higgs Signatures}}},
  \bibinfo{journal}{Phys.Rev.} \textbf{\bibinfo{volume}{D86}},
  \bibinfo{pages}{013009} (\bibinfo{year}{2012}), \eprint{1203.2924}.

\bibitem[{\citenamefont{Menon and Morrissey}(2009)}]{Menon:2009mz}
\bibinfo{author}{\bibfnamefont{A.}~\bibnamefont{Menon}} \bibnamefont{and}
  \bibinfo{author}{\bibfnamefont{D.~E.} \bibnamefont{Morrissey}},
  \emph{\bibinfo{title}{{Higgs Boson Signatures of MSSM Electroweak
  Baryogenesis}}}, \bibinfo{journal}{Phys. Rev.}
  \textbf{\bibinfo{volume}{D79}}, \bibinfo{pages}{115020}
  (\bibinfo{year}{2009}), \eprint{0903.3038}.

\bibitem[{\citenamefont{Batra et~al.}(2004)\citenamefont{Batra, Delgado,
  Kaplan, and Tait}}]{Batra:2003nj}
\bibinfo{author}{\bibfnamefont{P.}~\bibnamefont{Batra}},
  \bibinfo{author}{\bibfnamefont{A.}~\bibnamefont{Delgado}},
  \bibinfo{author}{\bibfnamefont{D.~E.} \bibnamefont{Kaplan}},
  \bibnamefont{and} \bibinfo{author}{\bibfnamefont{T.~M.~P.}
  \bibnamefont{Tait}}, \emph{\bibinfo{title}{{The Higgs mass bound in gauge
  extensions of the minimal supersymmetric standard model}}},
  \bibinfo{journal}{JHEP} \textbf{\bibinfo{volume}{02}}, \bibinfo{pages}{043}
  (\bibinfo{year}{2004}), \eprint{hep-ph/0309149}.

\bibitem[{\citenamefont{Maloney et~al.}(2006)\citenamefont{Maloney, Pierce, and
  Wacker}}]{Maloney:2004rc}
\bibinfo{author}{\bibfnamefont{A.}~\bibnamefont{Maloney}},
  \bibinfo{author}{\bibfnamefont{A.}~\bibnamefont{Pierce}}, \bibnamefont{and}
  \bibinfo{author}{\bibfnamefont{J.~G.} \bibnamefont{Wacker}},
  \emph{\bibinfo{title}{{D-terms, unification, and the Higgs mass}}},
  \bibinfo{journal}{JHEP} \textbf{\bibinfo{volume}{06}}, \bibinfo{pages}{034}
  (\bibinfo{year}{2006}), \eprint{hep-ph/0409127}.

\bibitem[{\citenamefont{Dine et~al.}(2007)\citenamefont{Dine, Seiberg, and
  Thomas}}]{Dine:2007xi}
\bibinfo{author}{\bibfnamefont{M.}~\bibnamefont{Dine}},
  \bibinfo{author}{\bibfnamefont{N.}~\bibnamefont{Seiberg}}, \bibnamefont{and}
  \bibinfo{author}{\bibfnamefont{S.}~\bibnamefont{Thomas}},
  \emph{\bibinfo{title}{{Higgs physics as a window beyond the MSSM (BMSSM)}}},
  \bibinfo{journal}{Phys. Rev.} \textbf{\bibinfo{volume}{D76}},
  \bibinfo{pages}{095004} (\bibinfo{year}{2007}), \eprint{0707.0005}.

\bibitem[{\citenamefont{Lu et~al.}(2014)\citenamefont{Lu, Murayama, Ruderman,
  and Tobioka}}]{Lu:2013cta}
\bibinfo{author}{\bibfnamefont{X.}~\bibnamefont{Lu}},
  \bibinfo{author}{\bibfnamefont{H.}~\bibnamefont{Murayama}},
  \bibinfo{author}{\bibfnamefont{J.~T.} \bibnamefont{Ruderman}},
  \bibnamefont{and} \bibinfo{author}{\bibfnamefont{K.}~\bibnamefont{Tobioka}},
  \emph{\bibinfo{title}{{A Natural Higgs Mass in Supersymmetry from
  NonDecoupling Effects}}}, \bibinfo{journal}{Phys.Rev.Lett.}
  \textbf{\bibinfo{volume}{112}}, \bibinfo{pages}{191803}
  (\bibinfo{year}{2014}), \eprint{1308.0792}.

\bibitem[{\citenamefont{Katz et~al.}(2014)\citenamefont{Katz, Reece, and
  Sajjad}}]{Katz:2014mba}
\bibinfo{author}{\bibfnamefont{A.}~\bibnamefont{Katz}},
  \bibinfo{author}{\bibfnamefont{M.}~\bibnamefont{Reece}}, \bibnamefont{and}
  \bibinfo{author}{\bibfnamefont{A.}~\bibnamefont{Sajjad}},
  \emph{\bibinfo{title}{{Naturalness, $b \to s \gamma$, and SUSY heavy
  Higgses}}}, \bibinfo{journal}{JHEP} \textbf{\bibinfo{volume}{10}},
  \bibinfo{pages}{102} (\bibinfo{year}{2014}), \eprint{1406.1172}.

\bibitem[{\citenamefont{Carena et~al.}(2013)\citenamefont{Carena, Nardini,
  Quiros, and Wagner}}]{Carena:2012np}
\bibinfo{author}{\bibfnamefont{M.}~\bibnamefont{Carena}},
  \bibinfo{author}{\bibfnamefont{G.}~\bibnamefont{Nardini}},
  \bibinfo{author}{\bibfnamefont{M.}~\bibnamefont{Quiros}}, \bibnamefont{and}
  \bibinfo{author}{\bibfnamefont{C.~E.} \bibnamefont{Wagner}},
  \emph{\bibinfo{title}{{MSSM Electroweak Baryogenesis and LHC Data}}},
  \bibinfo{journal}{JHEP} \textbf{\bibinfo{volume}{1302}}, \bibinfo{pages}{001}
  (\bibinfo{year}{2013}), \eprint{1207.6330}.

\bibitem[{\citenamefont{Profumo et~al.}(2007)\citenamefont{Profumo,
  Ramsey-Musolf, and Shaughnessy}}]{Profumo:2007wc}
\bibinfo{author}{\bibfnamefont{S.}~\bibnamefont{Profumo}},
  \bibinfo{author}{\bibfnamefont{M.~J.} \bibnamefont{Ramsey-Musolf}},
  \bibnamefont{and}
  \bibinfo{author}{\bibfnamefont{G.}~\bibnamefont{Shaughnessy}},
  \emph{\bibinfo{title}{{Singlet Higgs phenomenology and the electroweak phase
  transition}}}, \bibinfo{journal}{JHEP} \textbf{\bibinfo{volume}{08}},
  \bibinfo{pages}{010} (\bibinfo{year}{2007}), \eprint{0705.2425}.

\bibitem[{\citenamefont{Patel and Ramsey-Musolf}(2013)}]{Patel:2012pi}
\bibinfo{author}{\bibfnamefont{H.~H.} \bibnamefont{Patel}} \bibnamefont{and}
  \bibinfo{author}{\bibfnamefont{M.~J.} \bibnamefont{Ramsey-Musolf}},
  \emph{\bibinfo{title}{{Stepping Into Electroweak Symmetry Breaking: Phase
  Transitions and Higgs Phenomenology}}}, \bibinfo{journal}{Phys. Rev.}
  \textbf{\bibinfo{volume}{D88}}, \bibinfo{pages}{035013}
  (\bibinfo{year}{2013}), \eprint{1212.5652}.

\bibitem[{\citenamefont{Patel et~al.}(2013)\citenamefont{Patel, Ramsey-Musolf,
  and Wise}}]{Patel:2013zla}
\bibinfo{author}{\bibfnamefont{H.~H.} \bibnamefont{Patel}},
  \bibinfo{author}{\bibfnamefont{M.~J.} \bibnamefont{Ramsey-Musolf}},
  \bibnamefont{and} \bibinfo{author}{\bibfnamefont{M.~B.} \bibnamefont{Wise}},
  \emph{\bibinfo{title}{{Color Breaking in the Early Universe}}},
  \bibinfo{journal}{Phys. Rev.} \textbf{\bibinfo{volume}{D88}},
  \bibinfo{pages}{015003} (\bibinfo{year}{2013}), \eprint{1303.1140}.

\bibitem[{\citenamefont{Profumo et~al.}(2015)\citenamefont{Profumo,
  Ramsey-Musolf, Wainwright, and Winslow}}]{Profumo:2014opa}
\bibinfo{author}{\bibfnamefont{S.}~\bibnamefont{Profumo}},
  \bibinfo{author}{\bibfnamefont{M.~J.} \bibnamefont{Ramsey-Musolf}},
  \bibinfo{author}{\bibfnamefont{C.~L.} \bibnamefont{Wainwright}},
  \bibnamefont{and} \bibinfo{author}{\bibfnamefont{P.}~\bibnamefont{Winslow}},
  \emph{\bibinfo{title}{{Singlet-catalyzed electroweak phase transitions and
  precision Higgs boson studies}}}, \bibinfo{journal}{Phys. Rev.}
  \textbf{\bibinfo{volume}{D91}}, \bibinfo{pages}{035018}
  (\bibinfo{year}{2015}), \eprint{1407.5342}.

\bibitem[{\citenamefont{Jiang et~al.}(2015)\citenamefont{Jiang, Bian, Huang,
  and Shu}}]{Jiang:2015cwa}
\bibinfo{author}{\bibfnamefont{M.}~\bibnamefont{Jiang}},
  \bibinfo{author}{\bibfnamefont{L.}~\bibnamefont{Bian}},
  \bibinfo{author}{\bibfnamefont{W.}~\bibnamefont{Huang}}, \bibnamefont{and}
  \bibinfo{author}{\bibfnamefont{J.}~\bibnamefont{Shu}},
  \emph{\bibinfo{title}{{Impact of a complex singlet: From dark matter to
  baryogenesis}}} (\bibinfo{year}{2015}), \eprint{1502.07574}.

\bibitem[{\citenamefont{Blinov et~al.}(2015)\citenamefont{Blinov, Kozaczuk,
  Morrissey, and Tamarit}}]{Blinov:2015sna}
\bibinfo{author}{\bibfnamefont{N.}~\bibnamefont{Blinov}},
  \bibinfo{author}{\bibfnamefont{J.}~\bibnamefont{Kozaczuk}},
  \bibinfo{author}{\bibfnamefont{D.~E.} \bibnamefont{Morrissey}},
  \bibnamefont{and} \bibinfo{author}{\bibfnamefont{C.}~\bibnamefont{Tamarit}},
  \emph{\bibinfo{title}{{Electroweak Baryogenesis from Exotic Electroweak
  Symmetry Breaking}}} (\bibinfo{year}{2015}), \eprint{1504.05195}.

\bibitem[{\citenamefont{Inoue et~al.}(2015)\citenamefont{Inoue, Ovanesyan, and
  Ramsey-Musolf}}]{Inoue:2015pza}
\bibinfo{author}{\bibfnamefont{S.}~\bibnamefont{Inoue}},
  \bibinfo{author}{\bibfnamefont{G.}~\bibnamefont{Ovanesyan}},
  \bibnamefont{and} \bibinfo{author}{\bibfnamefont{M.~J.}
  \bibnamefont{Ramsey-Musolf}}, \emph{\bibinfo{title}{{Two-Step Electroweak
  Baryogenesis}}} (\bibinfo{year}{2015}), \eprint{1508.05404}.

\bibitem[{\citenamefont{Heinemeyer et~al.}(1998)\citenamefont{Heinemeyer,
  Hollik, and Weiglein}}]{Heinemeyer:1998kz}
\bibinfo{author}{\bibfnamefont{S.}~\bibnamefont{Heinemeyer}},
  \bibinfo{author}{\bibfnamefont{W.}~\bibnamefont{Hollik}}, \bibnamefont{and}
  \bibinfo{author}{\bibfnamefont{G.}~\bibnamefont{Weiglein}},
  \emph{\bibinfo{title}{{Precise prediction for the mass of the lightest Higgs
  boson in the MSSM}}}, \bibinfo{journal}{Phys. Lett.}
  \textbf{\bibinfo{volume}{B440}}, \bibinfo{pages}{296} (\bibinfo{year}{1998}),
  \eprint{hep-ph/9807423}.

\bibitem[{\citenamefont{Heinemeyer et~al.}(1999)\citenamefont{Heinemeyer,
  Hollik, and Weiglein}}]{Heinemeyer:1998np}
\bibinfo{author}{\bibfnamefont{S.}~\bibnamefont{Heinemeyer}},
  \bibinfo{author}{\bibfnamefont{W.}~\bibnamefont{Hollik}}, \bibnamefont{and}
  \bibinfo{author}{\bibfnamefont{G.}~\bibnamefont{Weiglein}},
  \emph{\bibinfo{title}{{The Masses of the neutral CP - even Higgs bosons in
  the MSSM: Accurate analysis at the two loop level}}}, \bibinfo{journal}{Eur.
  Phys. J.} \textbf{\bibinfo{volume}{C9}}, \bibinfo{pages}{343}
  (\bibinfo{year}{1999}), \eprint{hep-ph/9812472}.

\bibitem[{\citenamefont{Carena et~al.}(2000)\citenamefont{Carena, Haber,
  Heinemeyer, Hollik, Wagner, and Weiglein}}]{Carena:2000dp}
\bibinfo{author}{\bibfnamefont{M.}~\bibnamefont{Carena}},
  \bibinfo{author}{\bibfnamefont{H.~E.} \bibnamefont{Haber}},
  \bibinfo{author}{\bibfnamefont{S.}~\bibnamefont{Heinemeyer}},
  \bibinfo{author}{\bibfnamefont{W.}~\bibnamefont{Hollik}},
  \bibinfo{author}{\bibfnamefont{C.~E.~M.} \bibnamefont{Wagner}},
  \bibnamefont{and} \bibinfo{author}{\bibfnamefont{G.}~\bibnamefont{Weiglein}},
  \emph{\bibinfo{title}{{Reconciling the two loop diagrammatic and effective
  field theory computations of the mass of the lightest CP - even Higgs boson
  in the MSSM}}}, \bibinfo{journal}{Nucl. Phys.}
  \textbf{\bibinfo{volume}{B580}}, \bibinfo{pages}{29} (\bibinfo{year}{2000}),
  \eprint{hep-ph/0001002}.

\bibitem[{\citenamefont{Draper et~al.}(2012)\citenamefont{Draper, Meade, Reece,
  and Shih}}]{Draper:2011aa}
\bibinfo{author}{\bibfnamefont{P.}~\bibnamefont{Draper}},
  \bibinfo{author}{\bibfnamefont{P.}~\bibnamefont{Meade}},
  \bibinfo{author}{\bibfnamefont{M.}~\bibnamefont{Reece}}, \bibnamefont{and}
  \bibinfo{author}{\bibfnamefont{D.}~\bibnamefont{Shih}},
  \emph{\bibinfo{title}{{Implications of a 125 GeV Higgs for the MSSM and
  Low-Scale SUSY Breaking}}}, \bibinfo{journal}{Phys. Rev.}
  \textbf{\bibinfo{volume}{D85}}, \bibinfo{pages}{095007}
  (\bibinfo{year}{2012}), \eprint{1112.3068}.

\bibitem[{\citenamefont{Dorsch et~al.}(2013)\citenamefont{Dorsch, Huber, and
  No}}]{Dorsch:2013wja}
\bibinfo{author}{\bibfnamefont{G.~C.} \bibnamefont{Dorsch}},
  \bibinfo{author}{\bibfnamefont{S.~J.} \bibnamefont{Huber}}, \bibnamefont{and}
  \bibinfo{author}{\bibfnamefont{J.~M.} \bibnamefont{No}},
  \emph{\bibinfo{title}{{A strong electroweak phase transition in the 2HDM
  after LHC8}}}, \bibinfo{journal}{JHEP} \textbf{\bibinfo{volume}{10}},
  \bibinfo{pages}{029} (\bibinfo{year}{2013}), \eprint{1305.6610}.

\bibitem[{\citenamefont{Dorsch et~al.}(2014)\citenamefont{Dorsch, Huber,
  Mimasu, and No}}]{Dorsch:2014qja}
\bibinfo{author}{\bibfnamefont{G.~C.} \bibnamefont{Dorsch}},
  \bibinfo{author}{\bibfnamefont{S.~J.} \bibnamefont{Huber}},
  \bibinfo{author}{\bibfnamefont{K.}~\bibnamefont{Mimasu}}, \bibnamefont{and}
  \bibinfo{author}{\bibfnamefont{J.~M.} \bibnamefont{No}},
  \emph{\bibinfo{title}{{Echoes of the Electroweak Phase Transition:
  Discovering a second Higgs doublet through $A_0 \rightarrow ZH_0$}}},
  \bibinfo{journal}{Phys. Rev. Lett.} \textbf{\bibinfo{volume}{113}},
  \bibinfo{pages}{211802} (\bibinfo{year}{2014}), \eprint{1405.5537}.

\bibitem[{\citenamefont{Brust et~al.}(2012)\citenamefont{Brust, Katz, Lawrence,
  and Sundrum}}]{Brust:2011tb}
\bibinfo{author}{\bibfnamefont{C.}~\bibnamefont{Brust}},
  \bibinfo{author}{\bibfnamefont{A.}~\bibnamefont{Katz}},
  \bibinfo{author}{\bibfnamefont{S.}~\bibnamefont{Lawrence}}, \bibnamefont{and}
  \bibinfo{author}{\bibfnamefont{R.}~\bibnamefont{Sundrum}},
  \emph{\bibinfo{title}{{SUSY, the Third Generation and the LHC}}},
  \bibinfo{journal}{JHEP} \textbf{\bibinfo{volume}{03}}, \bibinfo{pages}{103}
  (\bibinfo{year}{2012}), \eprint{1110.6670}.

\bibitem[{\citenamefont{Aad et~al.}(2014{\natexlab{a}})}]{Aad:2014kra}
\bibinfo{author}{\bibfnamefont{G.}~\bibnamefont{Aad}} \bibnamefont{et~al.}
  (\bibinfo{collaboration}{ATLAS}), \emph{\bibinfo{title}{{Search for top
  squark pair production in final states with one isolated lepton, jets, and
  missing transverse momentum in $\sqrt s =$8 TeV $pp$ collisions with the
  ATLAS detector}}}, \bibinfo{journal}{JHEP} \textbf{\bibinfo{volume}{11}},
  \bibinfo{pages}{118} (\bibinfo{year}{2014}{\natexlab{a}}),
  \eprint{1407.0583}.

\bibitem[{CMS(2013{\natexlab{a}})}]{CMS-PAS-SUS-13-011}
\bibinfo{type}{Tech. Rep.} \bibinfo{number}{CMS-PAS-SUS-13-011},
  \bibinfo{institution}{CERN}, \bibinfo{address}{Geneva}
  (\bibinfo{year}{2013}{\natexlab{a}}),
  \urlprefix\url{http://cds.cern.ch/record/1547550}.

\bibitem[{CMS(2014)}]{CMS-PAS-SUS-13-009}
\bibinfo{type}{Tech. Rep.} \bibinfo{number}{CMS-PAS-SUS-13-009},
  \bibinfo{institution}{CERN}, \bibinfo{address}{Geneva}
  (\bibinfo{year}{2014}), \urlprefix\url{https://cds.cern.ch/record/1644584}.

\bibitem[{\citenamefont{Aad et~al.}(2014{\natexlab{b}})}]{Aad:2014nra}
\bibinfo{author}{\bibfnamefont{G.}~\bibnamefont{Aad}} \bibnamefont{et~al.}
  (\bibinfo{collaboration}{ATLAS}), \emph{\bibinfo{title}{{Search for
  pair-produced third-generation squarks decaying via charm quarks or in
  compressed supersymmetric scenarios in $pp$ collisions at $\sqrt{s}=8~$TeV
  with the ATLAS detector}}}, \bibinfo{journal}{Phys. Rev.}
  \textbf{\bibinfo{volume}{D90}}, \bibinfo{pages}{052008}
  (\bibinfo{year}{2014}{\natexlab{b}}), \eprint{1407.0608}.

\bibitem[{\citenamefont{Batell and Jung}(2015)}]{Batell:2015zla}
\bibinfo{author}{\bibfnamefont{B.}~\bibnamefont{Batell}} \bibnamefont{and}
  \bibinfo{author}{\bibfnamefont{S.}~\bibnamefont{Jung}},
  \emph{\bibinfo{title}{{Probing Light Stops with Stoponium}}},
  \bibinfo{journal}{JHEP} \textbf{\bibinfo{volume}{07}}, \bibinfo{pages}{061}
  (\bibinfo{year}{2015}), \eprint{1504.01740}.

\bibitem[{\citenamefont{Khachatryan et~al.}(2015)}]{Khachatryan:2014lpa}
\bibinfo{author}{\bibfnamefont{V.}~\bibnamefont{Khachatryan}}
  \bibnamefont{et~al.} (\bibinfo{collaboration}{CMS}),
  \emph{\bibinfo{title}{{Search for pair-produced resonances decaying to jet
  pairs in proton–proton collisions at $\sqrt{s}$=8 TeV}}},
  \bibinfo{journal}{Phys. Lett.} \textbf{\bibinfo{volume}{B747}},
  \bibinfo{pages}{98} (\bibinfo{year}{2015}), \eprint{1412.7706}.

\bibitem[{ATL(2015)}]{ATLAS-CONF-2015-026}
\bibinfo{type}{Tech. Rep.} \bibinfo{number}{ATLAS-CONF-2015-026},
  \bibinfo{institution}{CERN}, \bibinfo{address}{Geneva}
  (\bibinfo{year}{2015}), \urlprefix\url{http://cds.cern.ch/record/2037653}.

\bibitem[{\citenamefont{Drees et~al.}(1992)\citenamefont{Drees, Hagiwara, and
  Yamada}}]{Drees:1991zk}
\bibinfo{author}{\bibfnamefont{M.}~\bibnamefont{Drees}},
  \bibinfo{author}{\bibfnamefont{K.}~\bibnamefont{Hagiwara}}, \bibnamefont{and}
  \bibinfo{author}{\bibfnamefont{A.}~\bibnamefont{Yamada}},
  \emph{\bibinfo{title}{{Process independent radiative corrections in the
  minimal supersymmetric standard model}}}, \bibinfo{journal}{Phys. Rev.}
  \textbf{\bibinfo{volume}{D45}}, \bibinfo{pages}{1725} (\bibinfo{year}{1992}).

\bibitem[{\citenamefont{Blum and D'Agnolo}(2012)}]{Blum:2012kn}
\bibinfo{author}{\bibfnamefont{K.}~\bibnamefont{Blum}} \bibnamefont{and}
  \bibinfo{author}{\bibfnamefont{R.~T.} \bibnamefont{D'Agnolo}},
  \emph{\bibinfo{title}{{2 Higgs or not 2 Higgs}}}, \bibinfo{journal}{Phys.
  Lett.} \textbf{\bibinfo{volume}{B714}}, \bibinfo{pages}{66}
  (\bibinfo{year}{2012}), \eprint{1202.2364}.

\bibitem[{\citenamefont{Espinosa et~al.}(2012)\citenamefont{Espinosa, Grojean,
  Sanz, and Trott}}]{Espinosa:2012in}
\bibinfo{author}{\bibfnamefont{J.~R.} \bibnamefont{Espinosa}},
  \bibinfo{author}{\bibfnamefont{C.}~\bibnamefont{Grojean}},
  \bibinfo{author}{\bibfnamefont{V.}~\bibnamefont{Sanz}}, \bibnamefont{and}
  \bibinfo{author}{\bibfnamefont{M.}~\bibnamefont{Trott}},
  \emph{\bibinfo{title}{{NSUSY fits}}}, \bibinfo{journal}{JHEP}
  \textbf{\bibinfo{volume}{12}}, \bibinfo{pages}{077} (\bibinfo{year}{2012}),
  \eprint{1207.7355}.

\bibitem[{\citenamefont{Craig et~al.}(2015)\citenamefont{Craig, Farina,
  McCullough, and Perelstein}}]{Craig:2014una}
\bibinfo{author}{\bibfnamefont{N.}~\bibnamefont{Craig}},
  \bibinfo{author}{\bibfnamefont{M.}~\bibnamefont{Farina}},
  \bibinfo{author}{\bibfnamefont{M.}~\bibnamefont{McCullough}},
  \bibnamefont{and}
  \bibinfo{author}{\bibfnamefont{M.}~\bibnamefont{Perelstein}},
  \emph{\bibinfo{title}{{Precision Higgsstrahlung as a Probe of New Physics}}},
  \bibinfo{journal}{JHEP} \textbf{\bibinfo{volume}{03}}, \bibinfo{pages}{146}
  (\bibinfo{year}{2015}), \eprint{1411.0676}.

\bibitem[{\citenamefont{Fan et~al.}(2014)\citenamefont{Fan, Reece, and
  Wang}}]{Fan:2014axa}
\bibinfo{author}{\bibfnamefont{J.}~\bibnamefont{Fan}},
  \bibinfo{author}{\bibfnamefont{M.}~\bibnamefont{Reece}}, \bibnamefont{and}
  \bibinfo{author}{\bibfnamefont{L.-T.} \bibnamefont{Wang}},
  \emph{\bibinfo{title}{{Precision Natural SUSY at CEPC, FCC-ee, and ILC}}}
  (\bibinfo{year}{2014}), \eprint{1412.3107}.

\bibitem[{\citenamefont{Patel and Ramsey-Musolf}(2011)}]{Patel:2011th}
\bibinfo{author}{\bibfnamefont{H.~H.} \bibnamefont{Patel}} \bibnamefont{and}
  \bibinfo{author}{\bibfnamefont{M.~J.} \bibnamefont{Ramsey-Musolf}},
  \emph{\bibinfo{title}{{Baryon Washout, Electroweak Phase Transition, and
  Perturbation Theory}}}, \bibinfo{journal}{JHEP}
  \textbf{\bibinfo{volume}{07}}, \bibinfo{pages}{029} (\bibinfo{year}{2011}),
  \eprint{1101.4665}.

\bibitem[{\citenamefont{Cohen and Pierce}(2012)}]{Cohen:2011ap}
\bibinfo{author}{\bibfnamefont{T.}~\bibnamefont{Cohen}} \bibnamefont{and}
  \bibinfo{author}{\bibfnamefont{A.}~\bibnamefont{Pierce}},
  \emph{\bibinfo{title}{{Electroweak Baryogenesis and Colored Scalars}}},
  \bibinfo{journal}{Phys. Rev.} \textbf{\bibinfo{volume}{D85}},
  \bibinfo{pages}{033006} (\bibinfo{year}{2012}), \eprint{1110.0482}.

\bibitem[{\citenamefont{Moreno et~al.}(1997)\citenamefont{Moreno, Oaknin, and
  Quiros}}]{Moreno:1996zm}
\bibinfo{author}{\bibfnamefont{J.~M.} \bibnamefont{Moreno}},
  \bibinfo{author}{\bibfnamefont{D.~H.} \bibnamefont{Oaknin}},
  \bibnamefont{and} \bibinfo{author}{\bibfnamefont{M.}~\bibnamefont{Quiros}},
  \emph{\bibinfo{title}{{Sphalerons in the MSSM}}}, \bibinfo{journal}{Nucl.
  Phys.} \textbf{\bibinfo{volume}{B483}}, \bibinfo{pages}{267}
  (\bibinfo{year}{1997}), \eprint{hep-ph/9605387}.

\bibitem[{\citenamefont{Quiros}(1999)}]{Quiros:1999jp}
\bibinfo{author}{\bibfnamefont{M.}~\bibnamefont{Quiros}}, in
  \emph{\bibinfo{booktitle}{{High energy physics and cosmology. Proceedings,
  Summer School, Trieste, Italy, June 29-July 17, 1998}}}
  (\bibinfo{year}{1999}), pp. \bibinfo{pages}{187--259},
  \eprint{hep-ph/9901312},
  \urlprefix\url{http://alice.cern.ch/format/showfull?sysnb=0302087}.

\bibitem[{\citenamefont{Espinosa}(1996)}]{Espinosa:1996qw}
\bibinfo{author}{\bibfnamefont{J.}~\bibnamefont{Espinosa}},
  \emph{\bibinfo{title}{{Dominant two loop corrections to the MSSM finite
  temperature effective potential}}}, \bibinfo{journal}{Nucl.Phys.}
  \textbf{\bibinfo{volume}{B475}}, \bibinfo{pages}{273} (\bibinfo{year}{1996}),
  \eprint{hep-ph/9604320}.

\bibitem[{\citenamefont{Carena et~al.}(1998)\citenamefont{Carena, Quiros, and
  Wagner}}]{Carena:1997ki}
\bibinfo{author}{\bibfnamefont{M.~S.} \bibnamefont{Carena}},
  \bibinfo{author}{\bibfnamefont{M.}~\bibnamefont{Quiros}}, \bibnamefont{and}
  \bibinfo{author}{\bibfnamefont{C.}~\bibnamefont{Wagner}},
  \emph{\bibinfo{title}{{Electroweak baryogenesis and Higgs and stop searches
  at LEP and the Tevatron}}}, \bibinfo{journal}{Nucl.Phys.}
  \textbf{\bibinfo{volume}{B524}}, \bibinfo{pages}{3} (\bibinfo{year}{1998}),
  \eprint{hep-ph/9710401}.

\bibitem[{\citenamefont{Giardino et~al.}(2014)\citenamefont{Giardino, Kannike,
  Masina, Raidal, and Strumia}}]{Giardino:2013bma}
\bibinfo{author}{\bibfnamefont{P.~P.} \bibnamefont{Giardino}},
  \bibinfo{author}{\bibfnamefont{K.}~\bibnamefont{Kannike}},
  \bibinfo{author}{\bibfnamefont{I.}~\bibnamefont{Masina}},
  \bibinfo{author}{\bibfnamefont{M.}~\bibnamefont{Raidal}}, \bibnamefont{and}
  \bibinfo{author}{\bibfnamefont{A.}~\bibnamefont{Strumia}},
  \emph{\bibinfo{title}{{The universal Higgs fit}}}, \bibinfo{journal}{JHEP}
  \textbf{\bibinfo{volume}{1405}}, \bibinfo{pages}{046} (\bibinfo{year}{2014}),
  \eprint{1303.3570}.

\bibitem[{ATL(2013{\natexlab{a}})}]{ATLAS-CONF-2013-011}
\bibinfo{type}{Tech. Rep.} \bibinfo{number}{ATLAS-CONF-2013-011},
  \bibinfo{institution}{CERN}, \bibinfo{address}{Geneva}
  (\bibinfo{year}{2013}{\natexlab{a}}),
  \urlprefix\url{http://cds.cern.ch/record/1523696}.

\bibitem[{CMS(2013{\natexlab{b}})}]{CMS-PAS-HIG-13-018}
\bibinfo{type}{Tech. Rep.} \bibinfo{number}{CMS-PAS-HIG-13-018},
  \bibinfo{institution}{CERN}, \bibinfo{address}{Geneva}
  (\bibinfo{year}{2013}{\natexlab{b}}),
  \urlprefix\url{http://cds.cern.ch/record/1561758}.

\bibitem[{\citenamefont{Noble and Perelstein}(2008)}]{Noble:2007kk}
\bibinfo{author}{\bibfnamefont{A.}~\bibnamefont{Noble}} \bibnamefont{and}
  \bibinfo{author}{\bibfnamefont{M.}~\bibnamefont{Perelstein}},
  \emph{\bibinfo{title}{{Higgs self-coupling as a probe of electroweak phase
  transition}}}, \bibinfo{journal}{Phys.Rev.} \textbf{\bibinfo{volume}{D78}},
  \bibinfo{pages}{063518} (\bibinfo{year}{2008}), \eprint{0711.3018}.

\bibitem[{\citenamefont{Katz and Perelstein}(2014)}]{Katz:2014bha}
\bibinfo{author}{\bibfnamefont{A.}~\bibnamefont{Katz}} \bibnamefont{and}
  \bibinfo{author}{\bibfnamefont{M.}~\bibnamefont{Perelstein}},
  \emph{\bibinfo{title}{{Higgs Couplings and Electroweak Phase Transition}}},
  \bibinfo{journal}{JHEP} \textbf{\bibinfo{volume}{07}}, \bibinfo{pages}{108}
  (\bibinfo{year}{2014}), \eprint{1401.1827}.

\bibitem[{\citenamefont{Curtin et~al.}(2014)\citenamefont{Curtin, Meade, and
  Yu}}]{Curtin:2014jma}
\bibinfo{author}{\bibfnamefont{D.}~\bibnamefont{Curtin}},
  \bibinfo{author}{\bibfnamefont{P.}~\bibnamefont{Meade}}, \bibnamefont{and}
  \bibinfo{author}{\bibfnamefont{C.-T.} \bibnamefont{Yu}},
  \emph{\bibinfo{title}{{Testing Electroweak Baryogenesis with Future
  Colliders}}}, \bibinfo{journal}{JHEP} \textbf{\bibinfo{volume}{11}},
  \bibinfo{pages}{127} (\bibinfo{year}{2014}), \eprint{1409.0005}.

\bibitem[{ATL(2013{\natexlab{b}})}]{ATLAS-CONF-2013-012}
\bibinfo{type}{Tech. Rep.} \bibinfo{number}{ATLAS-CONF-2013-012},
  \bibinfo{institution}{CERN}, \bibinfo{address}{Geneva}
  (\bibinfo{year}{2013}{\natexlab{b}}),
  \urlprefix\url{http://cds.cern.ch/record/1523698}.

\bibitem[{\citenamefont{Khachatryan et~al.}(2014)}]{Khachatryan:2014ira}
\bibinfo{author}{\bibfnamefont{V.}~\bibnamefont{Khachatryan}}
  \bibnamefont{et~al.} (\bibinfo{collaboration}{CMS}),
  \emph{\bibinfo{title}{{Observation of the diphoton decay of the Higgs boson
  and measurement of its properties}}}, \bibinfo{journal}{Eur. Phys. J.}
  \textbf{\bibinfo{volume}{C74}}, \bibinfo{pages}{3076} (\bibinfo{year}{2014}),
  \eprint{1407.0558}.

\bibitem[{\citenamefont{Aad et~al.}(2013)}]{Aad:2013wqa}
\bibinfo{author}{\bibfnamefont{G.}~\bibnamefont{Aad}} \bibnamefont{et~al.}
  (\bibinfo{collaboration}{ATLAS}), \emph{\bibinfo{title}{{Measurements of
  Higgs boson production and couplings in diboson final states with the ATLAS
  detector at the LHC}}}, \bibinfo{journal}{Phys. Lett.}
  \textbf{\bibinfo{volume}{B726}}, \bibinfo{pages}{88} (\bibinfo{year}{2013}),
  \bibinfo{note}{[Erratum: Phys. Lett.B734,406(2014)]}, \eprint{1307.1427}.

\bibitem[{\citenamefont{Chatrchyan
  et~al.}(2014{\natexlab{a}})}]{Chatrchyan:2013mxa}
\bibinfo{author}{\bibfnamefont{S.}~\bibnamefont{Chatrchyan}}
  \bibnamefont{et~al.} (\bibinfo{collaboration}{CMS}),
  \emph{\bibinfo{title}{{Measurement of the properties of a Higgs boson in the
  four-lepton final state}}}, \bibinfo{journal}{Phys. Rev.}
  \textbf{\bibinfo{volume}{D89}}, \bibinfo{pages}{092007}
  (\bibinfo{year}{2014}{\natexlab{a}}), \eprint{1312.5353}.

\bibitem[{ATL(2013{\natexlab{c}})}]{ATLAS-CONF-2013-030}
\bibinfo{type}{Tech. Rep.} \bibinfo{number}{ATLAS-CONF-2013-030},
  \bibinfo{institution}{CERN}, \bibinfo{address}{Geneva}
  (\bibinfo{year}{2013}{\natexlab{c}}),
  \urlprefix\url{http://cds.cern.ch/record/1527126}.

\bibitem[{\citenamefont{Chatrchyan
  et~al.}(2014{\natexlab{b}})}]{Chatrchyan:2013iaa}
\bibinfo{author}{\bibfnamefont{S.}~\bibnamefont{Chatrchyan}}
  \bibnamefont{et~al.} (\bibinfo{collaboration}{CMS}),
  \emph{\bibinfo{title}{{Measurement of Higgs boson production and properties
  in the WW decay channel with leptonic final states}}},
  \bibinfo{journal}{JHEP} \textbf{\bibinfo{volume}{01}}, \bibinfo{pages}{096}
  (\bibinfo{year}{2014}{\natexlab{b}}), \eprint{1312.1129}.

\end{thebibliography}
\bibliographystyle{apsper}
\end{document}